\documentclass{article}

\usepackage{arxiv}

\usepackage[utf8]{inputenc} 
\usepackage[T1]{fontenc}    
\usepackage{hyperref}       
\usepackage{url}            
\usepackage{booktabs}       
\usepackage{amsfonts}       
\usepackage{nicefrac}       
\usepackage{microtype}      
\usepackage{lipsum}		
\usepackage{graphicx}

\usepackage{doi}
\usepackage{ulem}
\usepackage{amsmath}
\usepackage{float}
\usepackage{gensymb}
\usepackage{xcolor}
\usepackage{amsthm}
\usepackage{dcolumn}
\usepackage{epsfig}
\usepackage{bm}
\usepackage{datetime}
\usepackage[title]{appendix}
\usepackage{xr}

\makeatletter
\newcommand*{\addFileDependency}[1]{
\typeout{(#1)}
\@addtofilelist{#1}

\IfFileExists{#1}{}{\typeout{No file #1.}}
}\makeatother

\newcommand*{\myexternaldocument}[1]{%
\externaldocument{#1}
\addFileDependency{#1.tex}
}
\myexternaldocument{arxiv-si}
\usepackage[numbers,sort&compress]{natbib}
\bibpunct[, ]{[}{]}{,}{n}{,}{,}

\title{LeaPP: Learning Pathways to Polymorphs through machine learning analysis of atomic trajectories}

\author{ {\hspace{1mm}Steven W. Hall} \\
	Department of Chemical Engineering and Materials Science\\
	University of Minnesota\\
	Minneapolis, MN 55455 \\
	\And
	{\hspace{1mm}Porhouy Minh} \\
	Department of Chemistry\\
	University of Minnesota\\
	Minneapolis, MN 55455 \\
 	\And
	{\hspace{1mm}Sapna Sarupria} \\
	Department of Chemistry and\\
        Chemical Theory Center\\
	University of Minnesota\\
	Minneapolis, MN 55455 \\
	\texttt{sarupria@umn.edu} \\
}

\date{}

\renewcommand{\shorttitle}{LeaPP: Learning Pathways to Polymorphs}

\begin{document}
\maketitle

\begin{abstract}
Understanding the mechanisms underlying crystal formation is crucial. For most systems, crystallization typically goes through a nucleation process that involves dynamics that happen at short time and length scales. Due to this, molecular dynamics serves as a powerful tool to study this phenomenon. Existing approaches to study the mechanism often focus analysis on static snapshots of the global configuration, potentially overlooking subtle local fluctuations and history of the atoms involved in the formation of solid nuclei. To address this limitation, we propose a methodology that categorizes nucleation pathways into reactive pathways based on the time evolution of their constituent atoms. Our approach effectively captures the diverse structural pathways explored by crystallizing Lennard-Jones-like particles and solidifying Ni$_3$Al, providing a more nuanced understanding of nucleating pathways. Moreover, our methodology enables the prediction of the resulting polymorph from each reactive trajectory. This deep learning-assisted comprehensive analysis offers an alternative view of crystal nucleation mechanisms and pathways.
\end{abstract}

\section*{Introduction}

Crystal nucleation is an important phenomenon that has relevance in multiple scientific fields and technological applications such as biomineralization, pharmaceuticals, energy, and nanotechnology \cite{vekilov_nucleation_2010, MeldrumIMR2003, Sloan:book, muller_co2_2009, peysson_permeability_2012, Sosso_Crystal_2016}. The knowledge of the nucleation mechanisms and kinetics is of great interest for controlling crystallization relevant to these applications. For example, in the pharmaceutical industry, molecular understanding of protein and drug crystallization can guide strategies for polymorph and additive selections in drug design \cite{Sloan:book, vekilov_nucleation_2010, agarwal_solute_2014}.

In contrast to the macroscopically observable crystals that result from crystallization, the early critical stages of nucleation can involve as few as tens to hundreds of molecules. Accessing the short length- and timescales to analyze the fluctuations indicative of nucleation is difficult for experimental methods. Consequently, molecular simulations have emerged as an important tool in extracting mechanistic details of crystal nucleation and growth, though simulations have their challenges. Since nucleation is a rare event compared to the typical timescales of straightforward molecular simulations, direct sampling of nucleation is infeasible in most conditions that are not highly supersaturated. Many methods have been developed to circumvent this issue, such as path sampling \cite{Geissler:02:AdvChemPhys,Bolhuis:05:JComputPhys,tenWolde:06:JCP2,SarupriaJCP2022} or free energy surface exploration approaches \cite{LaioPNAS2002, RossoJCP2002}. Regardless of how nucleation is sampled, the discovery of mechanistically important structures and dynamics remains a time-consuming and challenging task, especially when studying a new system. This has motivated the development of methods to characterize the nucleation pathways and methods that find the types of structures and dynamics of nuclei that are predictive of nucleation and growth.

Correlating atomic coordinates from molecular simulations with structures important to nucleation requires knowledge of the quiescent liquid structure, as well as which deviations from the liquid signify crystal formation and growth. Often, these structures are in reference to bulk crystal phases, and order parameters have been developed to distinguish between bulk liquid and bulk crystalline environments. However, recognizing intermediate structures is less straightforward. For example, when applying well-established bond order parameters (BOPs) to a crystal nucleus surrounded by liquid, the resulting size can depend sensitively on the exact parameters of the BOP calculation \cite{Dellago:13:MolPhys,HajiAkbariJPCL2024}. Ideally, the local structures significant to nucleation would be detectable in a more generalized and unsupervised way. Several methods have been developed to advance the discovery of the variety of structures present in nucleating systems. Reinhart et al. identified and clustered the crystal structures of a nucleating colloidal system by comparing the local atomic neighborhood graphs \cite{PanagiotopoulosSM2017}. Previous work with an autoencoder using Steinhardt BOPs \cite{SteinhardtPRB1983} has found that the local environments during nucleation can be reasonably organized in a low-dimensional space in an unsupervised manner \cite{FilionJCP2019}. Many other methods have also followed this general schema of combining machine learning (ML) approaches with local structural descriptors to organize and cluster the atomic environments observed in crystallizing systems \cite{Glotzer:20:JPCB,MittalSoftMatt2021,ReinhartCMS2021,LiJCP2022,Dschemuchadse2024MRSA}.

Alternatively to the aforementioned analyses of local structure, the short-time dynamics has also been used to describe and differentiate local atomic environments. Dynamical propensity has been introduced as a measure of the mobility of an atom based on its local structure \cite{MarcoJPCB2014}. A decrease in dynamical propensity has been detected as a precursor to ice \cite{MichaelidesPNAS2019} and Ni nucleation \cite{RogalFaraday2022}. Ishiai et al. used a graph neural network with contrastive learning to learn atomic latent space coordinates \cite{YasuokaJCTC2024}. The contrastive loss sought to minimize the distances between subsequent frames of atom neighborhood graph representations. The local environments and neighbors shuffling (LENS) method seeks to define dynamical groups of atoms based on how rapidly the indices of neighboring atoms change \cite{PavanPNAS2023}.

Nucleation is a collective process involving many atoms and can proceed through a variety of structures. While the previously mentioned methods can help discover these pathways and learn useful local representations, the similarities and differences between the time evolution of the involved atoms may hold additional information on the nucleation process. Here, we develop a methodology to group nucleation trajectories into pathways based on the time evolution of their constituent atoms, which we call LeaPP: Learning Pathways to Polymorphs. Our method relies on machine learning-based dimensionality reduction to learn meaningful atomic representations from generalized local features. The atoms involved in nucleation are then clustered based on the time evolution of their representation. We demonstrate our method on two example systems---crystallization of particles with the 7--6 potential (a modified Lennard-Jones potential with softer interactions) and crystallization of Ni$_3$Al from the melt. We find that our method is capable of resolving the variety of structural pathways sampled by these systems and groups the nucleation trajectories into clusters based on the crystalline phase composition of the nuclei. We observe that the resulting clusters of nucleation trajectories are predictive of the growth phase polymorphism. We expect that such a methodology that considers the time evolution of the building blocks will prove useful to a wide range of self-assembly problems.

\section*{LeaPP: Learning Pathways to Polymorphs}
\begin{figure*} 
\centering
\includegraphics[width=\linewidth]{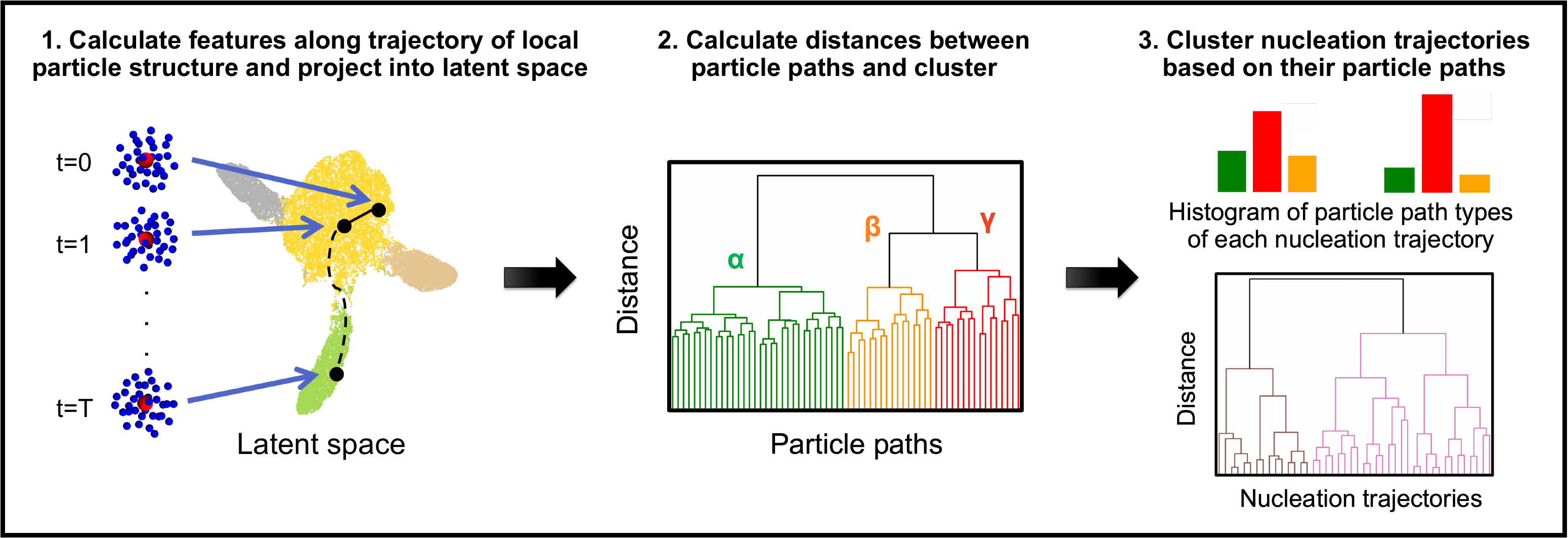}
\caption{Overview of LeaPP. The time trajectory of the local environment of each particle is embedded into a low-dimensional latent space for a set of nucleation trajectories. The pairwise distances between particle paths in the latent space are then calculated and clustered to give particle path clusters. Each nucleation trajectory is then characterized by its distribution of particle path clusters. These distributions are used to calculate the pairwise distances between nucleation trajectories and cluster them into distinct pathways.}
\label{fig:overview}
\end{figure*}

We develop a method called LeaPP, that accounts for the particle path time evolution to characterize nucleation pathways. We demonstrate that this approach provides a richer insight into the nucleation mechanism than that obtained by the traditional approach that focuses generally only on structural analysis per frame.

LeaPP consists of three key steps, as outlined schematically in Fig.~\ref{fig:overview}. In the first step, the particles relevant to nucleation are identified, and their local environments are characterized by a set of local descriptors that take the relative positions of nearest neighbors as input. The selected particles are those that are identified as having ever been a part of the largest solid nucleus at any frame in the simulation. The set of descriptors is comprised of Steinhardt order parameters $q_l$ \cite{SteinhardtPRB1983} as well as their neighbor-averaged variants $\overline{q}_l$ \cite{Dellago:08:JCP}. The local order parameters are calculated for $l=\{2,4,6,8\}$ to keep the features general. For each $l$, multiple cutoff distances for the consideration of nearest neighbors are used. A variational autoencoder (VAE) is then trained to generate a low-dimensional representation of each particle's local environment (details in Materials and Methods). The time trajectories of the particles are projected into the latent space, giving a set of particle paths from the nucleation trajectories.

In the second step, the particle paths in the latent space are clustered using hierarchical agglomerative clustering based on the pairwise distances. The pairwise distances between particle paths are calculated using dynamic time warping (DTW) \cite{ChibaITASS1978,KeoghDMKD2017}. In Fig.~\ref{fig:overview}, these clusters are labeled $\alpha$, $\beta$, and $\gamma$ as an example. These particle path clusters serve as our labels for nucleation particles, from which we can extract further information about their role in nucleation. As an example, we demonstrate this concept on a direct coexistence trajectory between liquid and solid particles that interact via the 7--6 potential. The result shows that LeaPP particle path clusters (I, II, and III) correspond to the liquid, solid, and interfacial particles, respectively (see \textit{SI Appendix}, Fig.~S2). This result effectively shows that LeaPP learns to differentiate particles in an unsupervised manner via the evolution of particles local environment. Similarly, we seek to demonstrate this learning ability from LeaPP in our nucleation systems as well.

In step three, we characterize the nucleation trajectories into groups of nucleation pathways. To perform this characterization, first, we construct a normalized histogram for each nucleation trajectory. The histogram counts the total number of different particle types (i.e., $\alpha$, $\beta$, and $\gamma$ in Fig.~\ref{fig:overview}) in the largest solid nucleus throughout the frames of a trajectory. For example, if a particle labeled $\alpha$ is present in the nucleus in frame 1 and frame 10 of the nucleation trajectory, this particle has contributed a total count of 2 towards the $\alpha$ particles in the (unnormalized) histogram. Once each nucleation trajectory has a histogram, we compute the pairwise differences between all nucleation trajectory histograms and obtain the difference metric for hierarchical agglomerative clustering of the nucleation trajectories (Step 3 of Fig.~\ref{fig:overview}). The resulting clusters from this last step will constitute the distinct nucleation pathways of the system. More details of the LeaPP method are present in Materials and Methods.

\section*{7--6 Crystal Nucleation}
We first illustrate LeaPP with a system of particles interacting with the 7--6 potential \cite{Shi_Structure_2011}. 7--6 potential is a modified Lennard-Jones potential with softer attractive and repulsive forces near the potential minimum. Nucleation trajectories were obtained with path sampling simulations. Simulation and path sampling details are in Materials and Methods.

\subsection*{Validation of VAE projection} 

\begin{figure*} 
\centering
\includegraphics[width=0.9\textwidth]{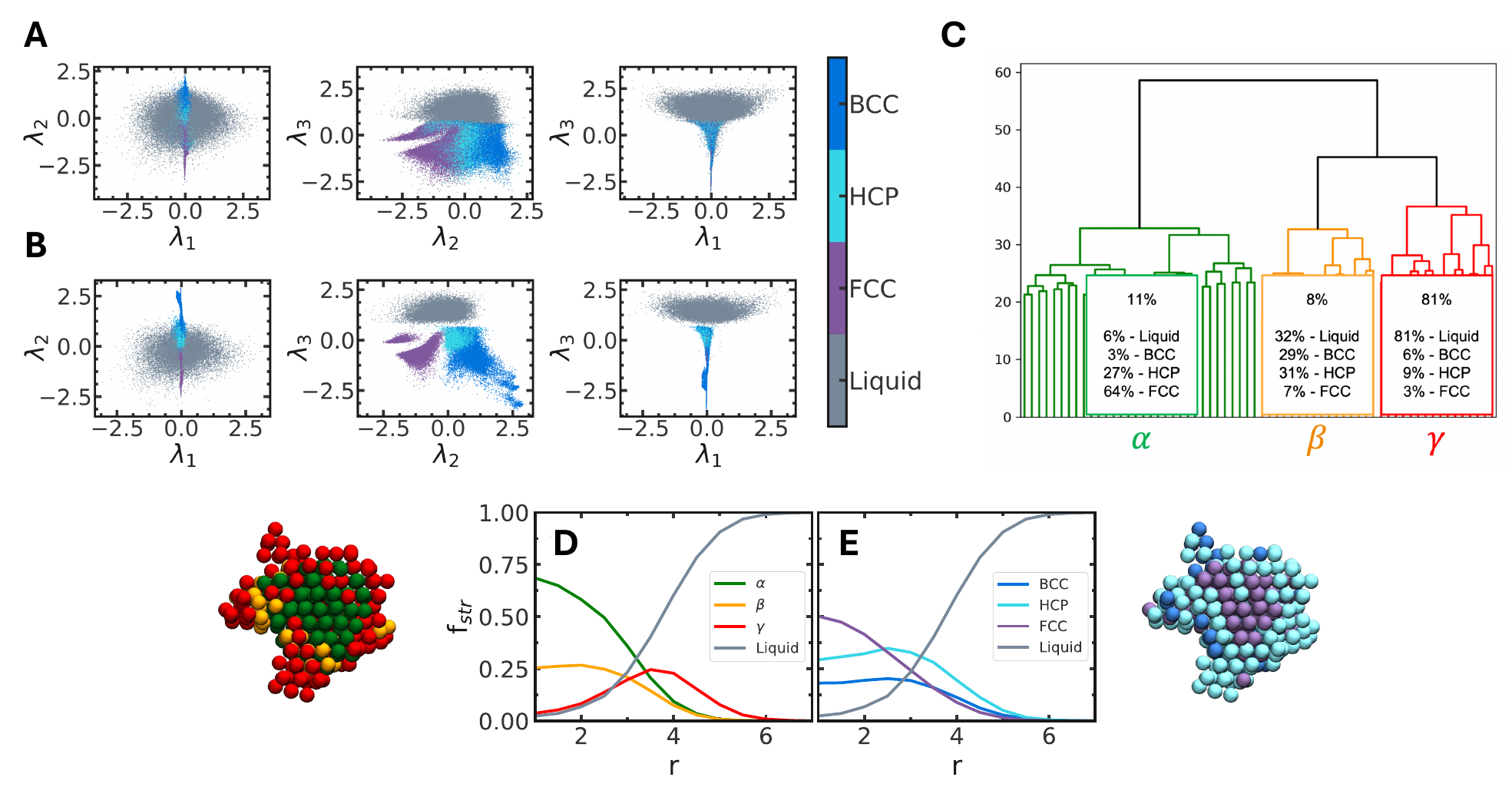}
\caption{\textit{(A)} Latent space projection of the 7--6 training data with phase labels. Each panel represents a 2D plane of the 3D latent space containing the training data colored by phase labels. Labels are acquired based on the values of $\overline{q}_{4}$ and $\overline{q}_{6}$. \textit{(B)} Latent space projection of the 7--6 particles from simulations of bulk phases. \textit{(C)} Particle path clusters for 7--6 particles. The clustering protocol is detailed in Materials and Methods section. Each colored region of the dendrogram represents a cluster of particle paths ($\alpha$, $\beta$, $\gamma$). The percentage at the top center is the fraction of particle paths that belong to each cluster. The percentage of liquid, BCC, FCC, and HCP is calculated from the label of the particle in the last frame. \textit{(D-E)} Radial composition analysis of configurations from 181 7--6 nucleation trajectories. $r$ is the distance from the nucleus center of mass, and f$_\text{str}$ is the fraction of each type of particle calculated at each spherical shell. In panel D, the radial composition is calculated using the particle path cluster labels ($\alpha$, $\beta$, $\gamma$) from panel B. The same calculation is performed for panel E with labels from their $\overline{q}_4$ and $\overline{q}_6$ values. Any particles that do not belong to the solid nucleus are labeled liquid. The analysis is performed using one configuration from each nucleation trajectory with an estimated committor of 0.9 or greater. }
\label{fig:lj-vae-pp3-rcaoverall}
\end{figure*}

Following the training of the VAE, the latent space is evaluated. The distribution of training samples within the latent space is presented in Fig.~\ref{fig:lj-vae-pp3-rcaoverall}. Labels are assigned based on the values of $\overline{q}_{4}$ and $\overline{q}_{6}$. Specifically, the identification of liquid and solid phases (FCC, BCC, and HCP) is established by categorizing particles with $\overline{q}_6 < 0.36$ as liquid. Further categorization of solid phases is based $\overline{q}_4$, where BCC corresponds to $\overline{q}_4 < 0.07$, HCP to $0.07 \leq \overline{q}_4 < 0.12$, and FCC to $\overline{q}_4 \geq 0.12$. The distinction between liquid and solid particles is evident, whereas FCC, BCC, and HCP particles exhibit some overlap within the latent space.

Since the labels of the projections in Fig.~\ref{fig:lj-vae-pp3-rcaoverall}A are obtained from $\overline{q}_{4}$ and $\overline{q}_{6}$, this renders the method reliant upon traditional BOPs to evaluate the performance of the method. To reduce this reliance, we project samples from simulated bulk phases (liquid, FCC, BCC, HCP) onto the latent space using the same VAE architecture (see Fig. \ref{fig:lj-vae-pp3-rcaoverall}B). This projection reaffirms the clear differentiation between liquid and solid particles and a distinct separation between FCC and BCC/HCP. However, there is some overlap between the BCC and HCP projections. Collectively, this indicates that the VAE is learning the local environmental nuances observed in nucleation trajectories.

\subsection*{7--6 particle path clustering} 

After validating the information learned by the model, particle paths are projected onto the latent space and clustered according to the similarity in their time evolution (Step 2 of LeaPP). In essence, particles with similar local environments that also evolve similarly are clustered together. To develop an understanding of the clusters, we label them based on their $\overline{q}_4$ and $\overline{q}_6$ values in the last frame (in basin B). $\overline{q}_4$ and $\overline{q}_6$ are selected based on their ability to distinguish the bulk particle phases. Indeed, other approaches to understand the clusters are possible. For example, we could base the interpretation on the composition of particles in all frames during nucleation. However, this leads to an overwhelming prevalence of liquid particles in each cluster since the initial frames of the nucleation trajectory predominantly consist of liquid particles rather than solid ones (see \textit{SI Appendix}, Fig.~S3). 

The distance cutoff for determining the number of clusters is selected to ensure that the composition of each cluster is distinct and that each particle path cluster represents a statistically relevant percentage of particle paths. For instance, if we use a cutoff of 30 in Fig. \ref{fig:lj-vae-pp3-rcaoverall}C, this results in nine particle path clusters, but analysis reveals that their compositions are not distinct. Thus, a higher cutoff is chosen to yield three distinct clusters---referred to as $\alpha$, $\beta$, and $\gamma$. Notably, the majority of particles are associated with the $\gamma$ cluster, followed by the $\alpha$ cluster, and, lastly, the $\beta$ cluster. The $\gamma$ cluster is predominantly composed of liquid particles, with a minority of BCC, HCP, and FCC particles. To better understand the reason behind this composition, a reexamination of our particle selection criteria is warranted. Our particle selection criterion only necessitates a particle to be part of the largest nucleus in at least one frame. Consequently, the overall count of liquid particles sampled throughout the trajectories outnumbers that of solids. Given that the $\gamma$ cluster represents the largest percentage of particles, this observation logically aligns with our particle selection criteria, and the $\gamma$ cluster exhibits a high percentage of liquid in the last frame. We also find that $\gamma$ particles correlate to "cloud particles" introduced by Lechner et al. \cite{Lechner_Role_2011} in the $\overline{q}_4 $ and $ \overline{q}_6$ space, where these particles are more structured than liquids and less structured than bulk solids (see \textit{SI Appendix}, Fig.~S4). Since many of the particles in this cluster end in the liquid phase, this suggests that the $\gamma$ cluster represents particles that are only transiently part of the solid nucleus. In contrast, the $\alpha$ cluster is predominantly comprised of FCC particles, with some HCP particles and a few liquid and BCC particles. Meanwhile, the $\beta$ cluster displays a relatively balanced distribution of liquid, BCC, and HCP atoms, with a sparse presence of FCC atoms. These two particle path clusters, representing less than one-third of the tracked particles, exhibit a higher percentage of solids in their definition at the last frame.

To further understand the connection between the particle path clusters and the nucleation mechanism, we performed a radial composition analysis (RCA) of configurations obtained from the nucleation trajectories. We selected one configuration with a committor probability of 0.9 or greater per nucleation trajectory for the RCA. This analysis describes the composition of the largest solid nucleus of the configuration as a function of the distance from the nucleus center of mass. RCA results reveal that particle path clusters provide valuable information about the spatial distribution of particles within the solid nucleus (Fig. \ref{fig:lj-vae-pp3-rcaoverall}D and E). Specifically, particles belonging to the $\alpha$ and $\beta$ clusters correlate with the core particles of the solid nucleus, while the $\gamma$ cluster is associated with interfacial particles. These insights are attributed to the well-separated latent space between liquids and solids, as well as our classification of particles through their time evolution. By capturing the time evolution, we effectively distinguish between core and interfacial particles within the solid nucleus. The evolving trajectories within the $\gamma$ cluster likely indicate fluctuations in and out of the liquid phase in the latent space, consistent with their role at the nucleus interface. Consequently, particles from the $\alpha$ and $\beta$ clusters emerge as comprising the core of their respective solid nuclei, which may exhibit various crystal structures such as HCP, BCC, or FCC. 

The discovery of this spatial characterization of the nucleus opens up new avenues for studying and characterizing interfacial particles through correlations of various structural and dynamical properties to this region. This marks a notable improvement from traditional order parameters, as this method does not rely on bulk structural definitions to identify interfacial structures, which often exhibit less bulk-like characteristics and instead possess more quasi-solid qualities.

\subsection*{7--6 nucleation trajectory clustering} 

\begin{figure*}
\centering
\includegraphics[width=\textwidth]{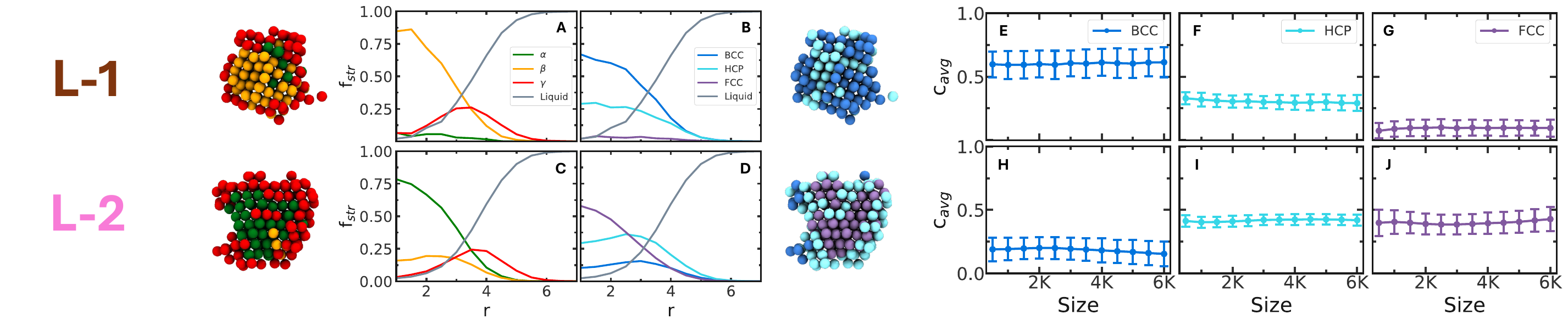}
\caption{\textit{(A-D)} Radial composition analysis of 7--6 configurations from each trajectory cluster in \textit{SI Appendix}, Fig.~S5. Panel A and C use the particle path cluster labels from Fig. \ref{fig:lj-vae-pp3-rcaoverall}B for the analysis, whereas panel B and D use  $\overline{q}_4$ and $\overline{q}_6$ values. The radial compositions are computed for (A-B) L-1 cluster and (C-D) L-2 cluster of trajectories from \textit{SI Appendix}, Fig.~S5. Configurations with committor estimates greater than or equal to 0.9 are again selected for the calculation. \textit{(E-J)} Average composition of 7--6 nucleation trajectories from each trajectory cluster in \textit{SI Appendix}, Fig.~S5. The x-axis represents the average nucleus size of the configurations in the trajectories of each respective cluster. c$_\text{avg}$ represents the average composition of the nucleus at a specific size according to the three phases: BCC, HCP, and FCC. Blue lines represent the BCC composition, cyan lines represent the HCP composition and purple lines represent the FCC composition. The composition is calculated for trajectories in L-1 cluster (E-G) and trajectories in L-2 cluster (H-J). The error bars reflect the standard deviation of the average composition across all trajectories launched within each trajectory cluster.}
\label{fig:lj-trajscls-rca-2retcls}
\end{figure*}

 We use the composition of particle path types in the nucleus to cluster the nucleation trajectories and characterize different nucleation pathways. Our approach, detailed in Materials and Methods, involves constructing a histogram of particle path types for each nucleation trajectory and comparing the trajectory histograms in a pairwise manner to calculate a distance matrix (Step 3 of LeaPP). The distance matrix is used for hierarchical agglomerative clustering, which yields clusters of nucleation trajectories. Our decision to calculate nucleation trajectory differences using histograms stems from initial visualizations of the trajectories. We observed variations in the distribution of $\alpha$ and $\beta$ particles across different trajectories. Some trajectories showed a predominance of $\alpha$ particles throughout, while others exhibited a higher proportion of $\beta$ particles. Inspired by these observations, we adopted histograms to capture the relative abundance of each particle path type in the nucleation trajectory. 

With a complete linkage cutoff distance of 1.2 for hierarchical clustering, the nucleation trajectories are separated into two clusters: L-1 and L-2 (\textit{SI Appendix}, Fig.~S5). To further probe the two trajectory clusters (L-1 and L-2), an additional RCA analysis is performed on each of them (Fig. \ref{fig:lj-trajscls-rca-2retcls}). RCA results reveal that these two nucleation trajectories are distinguished via the difference in their nucleus cores. Configurations of L-1 nucleation trajectories (Fig. \ref{fig:lj-trajscls-rca-2retcls}A and \ref{fig:lj-trajscls-rca-2retcls}B) have a dominant $\beta$ core, with $\gamma$ interfacial structures. In the $\overline{q}_4 $ and $ \overline{q}_6$ interpretation, these configurations are rich in BCC content in the core with some HCP and have low FCC content overall. As for L-2 nucleation trajectories (Fig. \ref{fig:lj-trajscls-rca-2retcls}C and \ref{fig:lj-trajscls-rca-2retcls}D), these configurations have a dominant $\alpha$ core, a small amount of $\beta$ particles near the nucleus interfacial layer, and $\gamma$ interfacial particles. In the $\overline{q}_4 $ and $ \overline{q}_6$ interpretation, these configurations are rich in FCC content in the core, HCP content on the interface, and low BCC content overall. 

Given the radial composition profile of L-1 and L-2 observed in Fig. \ref{fig:lj-trajscls-rca-2retcls}, we explore the possibility of polymorph prediction via trajectory clustering. Five trajectories are extended from each trajectory's last configuration until the solid nucleus reaches 6000 particles in size to confirm the existence of the two polymorphs. Each set of five trajectories is averaged to represent the average composition of an extended trajectory from each configuration. Then all trajectories from each configuration are averaged once more to give the representative trajectory of each trajectory cluster. Figure \ref{fig:lj-trajscls-rca-2retcls}E-J demonstrates that LeaPP allows for polymorph prediction via trajectory clustering. Trajectories extending from L-1 nucleation trajectories (Fig. \ref{fig:lj-trajscls-rca-2retcls}) maintain a higher BCC content compared to the other phases. These trajectories also have a low fraction of FCC. As such, this trajectory cluster is dubbed the "BCC-dominant" trajectory cluster. Trajectories extending from L-2 nucleation trajectories (Fig. \ref{fig:lj-trajscls-rca-2retcls}) present an FCC-HCP dominance with a low fraction of BCC. Thus, these trajectories are dubbed as "FCC-HCP dominant."   

Although the choice of the inter-cluster cutoff distance for trajectory clustering allows us to see two clear polymorphs, decreasing the cutoff, thus increasing the number of nucleation trajectory clusters, results in three trajectory clusters: L-1, L-2.1, and L-2.2 (see \textit{SI Appendix}, Fig.~S6-S8). From the RCA results at post-nucleation size (\textit{SI Appendix}, Fig.~S7), L-1 cluster's configurations maintain the same composition as shown in \textit{SI Appendix}, Fig.~S7A. L-2.1 cluster's configurations have a dominant $\alpha$ core with $\beta$ particles near the nucleus interfacial layer and $\gamma$ interfacial particles (\textit{SI Appendix}, Fig.~S7C). Lastly, L-2.2 cluster's configurations have a mixture of the $\alpha$ and $\beta$ core with $\gamma$ interfacial particles (\textit{SI Appendix}, Fig.~S7E). In the $\overline{q}_4 $ and $ \overline{q}_6$ interpretation, configurations of L-1 have a BCC-dominant core with small amounts of HCP, configurations of L-2.1 have a dominant FCC core with HCP interface, and configurations from L-2.2 have a mixture of all the phases throughout the cluster, with HCP having the highest fraction, followed by BCC and FCC, respectively. 

After extending the trajectories into the growth phase, we observe that only two polymorph profiles exist (\textit{SI Appendix}, Fig.~S8). The composition of trajectories from L-2.2 is not distinguishable from L-2.1 due to the high uncertainty. Although the composition of trajectories from this new cluster does not suggest that it introduces any new polymorph, the sensitivity of LeaPP to the relative fraction of $\alpha$ and $\beta$ particles in the nucleus has allowed us to tease out two different types of nucleus composition. 

\subsection*{Benchmarking against traditional approach}

Traditionally, the structures of nuclei (and polymorphs) in nucleating trajectories from molecular simulations rely on structure-based order parameters calculated on a per-configuration basis. For example, in systems similar to the 7--6 system, the Steinhardt order parameters that discern the presence of crystal structures such as BCC, FCC, and HCP during nucleation are used. However, this method may not always be reliable when dealing with the dynamic nature of nucleation trajectories that grow into and out of (metastable) structures. This could result in a similar classification of nucleation trajectories with subtle differences in the composition fluctuations within a nucleus, thus, possibly overlooking some nucleation pathways. In light of this, we compare the potential of polymorph prediction based on Steinhardt order parameters in the 7-6 system with the predictions from LeaPP.

\begin{figure}
    \centering
    \includegraphics[width=\textwidth]{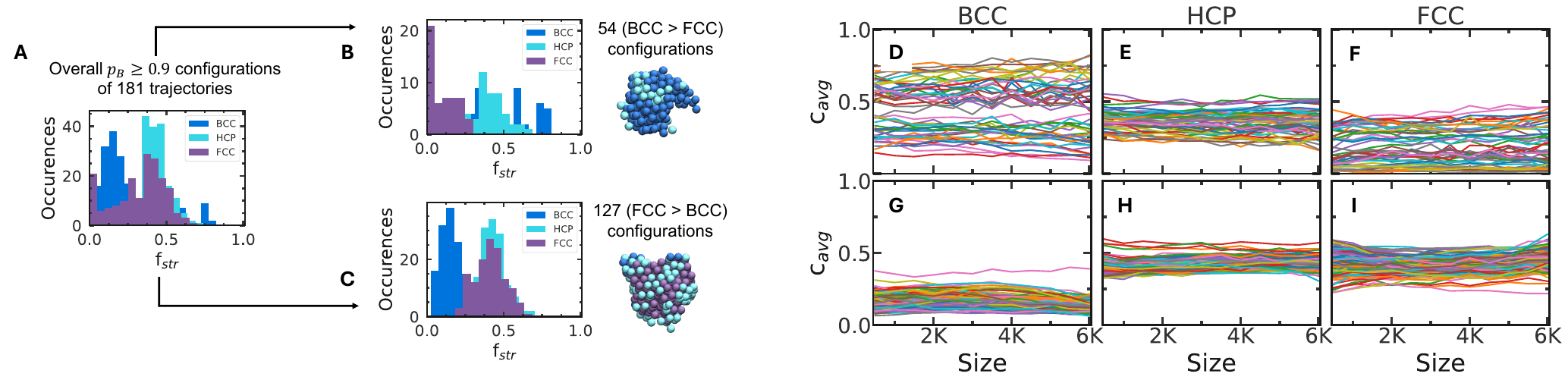}
    \caption{\textit{(A-C)} Configuration selection for composition-based clustering of 7--6 nucleation trajectories. The y-axis reflects the number of configurations with the BCC/HCP/FCC fraction indicated on the x-axis. Color code: Blue -- BCC, cyan -- HCP, and purple -- FCC. Panel A shows the composition distribution of all 7--6 nucleation trajectories' last configuration. Panel B and C, respectively, show a distribution of configurations that have a higher fraction of BCC than FCC, and a distribution of configurations that have a higher fraction of FCC than BCC. \textit{(D-I)} Average composition of extended trajectories from "BCC$>$FCC" and "FCC$>$BCC" configurations. The x-axis represents the average nucleus size of the configurations in the trajectories of each respective cluster. c$_\text{avg}$ represents the average composition of the nucleus at a specific size according to the three phases: BCC (D,G), HCP (E,H) and FCC (F,I). Each line in the plot represents the average composition of five trajectories extended from each "BCC$>$FCC" configuration (D-F) or "FCC$>$BCC" configuration (G-I).}
    \label{fig:lj-compare-methods-q4q6}
\end{figure}

Figure \ref{fig:lj-compare-methods-q4q6}D-I demonstrates the result of composition-based clustering of 7--6 nucleation trajectories based on the $\overline{q}_{4}$ and $\overline{q}_{6}$ interpretation. To benchmark this method against LeaPP, we choose a clustering method on the same 7--6 nucleation trajectories that would also give us two clusters of trajectories. Based on the bimodal composition distribution of all 7--6 trajectories' last configuration in Fig. \ref{fig:lj-compare-methods-q4q6}A, the two clusters are (i) trajectories that are extended from configurations where the fraction of BCC is larger than FCC (BCC$>$FCC) in Fig. \ref{fig:lj-compare-methods-q4q6}B, and (ii) trajectories that are extended from configurations where the fraction of FCC is larger than BCC (FCC$>$BCC) in Fig. \ref{fig:lj-compare-methods-q4q6}C. 

\begin{figure}[H]
\centering
\includegraphics[width=0.6\linewidth]{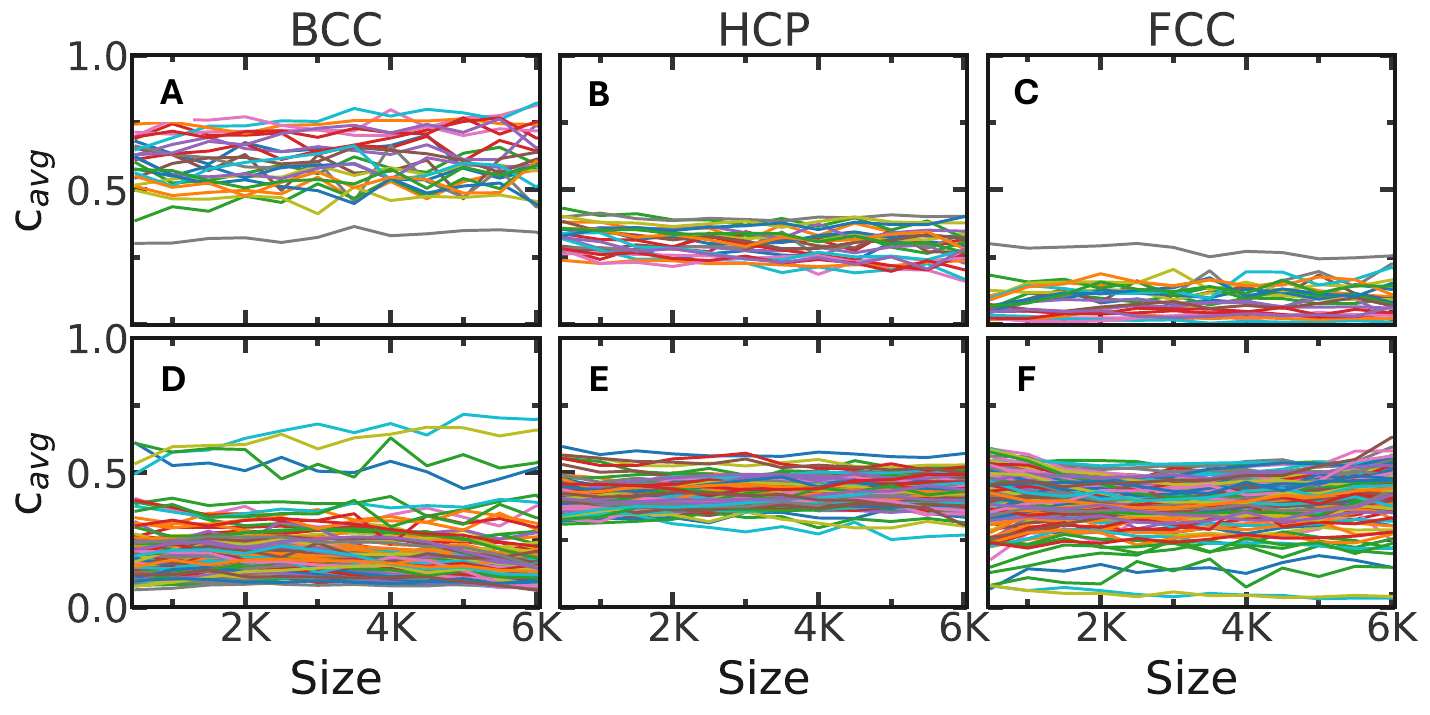}
\caption{Average composition of extended trajectories from two nucleation trajectory clusters found via LeaPP in \textit{SI Appendix}, Fig.~S5 of the 7--6 system. Each line represents the average composition of five trajectories extended from configurations of L-1 trajectory cluster (A-C) or L-2 trajectory cluster (D-F).}
\label{fig:lj-compare-methods-MLresult}
\end{figure}

Based on this criteria, trajectories that are extended from "FCC$>$BCC" configurations correlate well with the FCC-HCP dominant trajectories found in Fig. \ref{fig:lj-trajscls-rca-2retcls}H-J. These trajectories tend to maintain a balanced fraction of FCC and HCP, which is generally higher than the fraction of BCC. Whereas trajectories extended from "BCC$>$FCC" configurations do not seem to have a clear polymorph profile. This cluster includes trajectories that can be high or low in the BCC fraction (low FCC). This demonstrates that clustering trajectories and ultimately determining the polymorph via the static picture of a nucleus composition is insufficient. Configurations can grow and quickly change their composition from where they start. In this case, BCC configurations can slightly melt or grow to form more stable structures like HCP and FCC during their growth. LeaPP, which accounts for the evolution of particle paths in the growing nucleus of each nucleation trajectory, can cluster the nucleation trajectories more accurately, reflecting two polymorph profiles in the trajectories (Fig. \ref{fig:lj-compare-methods-MLresult}).

\section*{Ni$_3$Al Crystal Nucleation}
To demonstrate LeaPP's versatility in studying more realistic nucleation systems, we apply this method to perform polymorph prediction on Ni$_{3}$Al nucleation from the melt. This system is shown to exhibit two different polymorph profiles in previous work by Liang et al.\cite{Liang2020}.

\subsection*{Validation of VAE projection}

\begin{figure*}
\centering
\includegraphics[width=\textwidth]{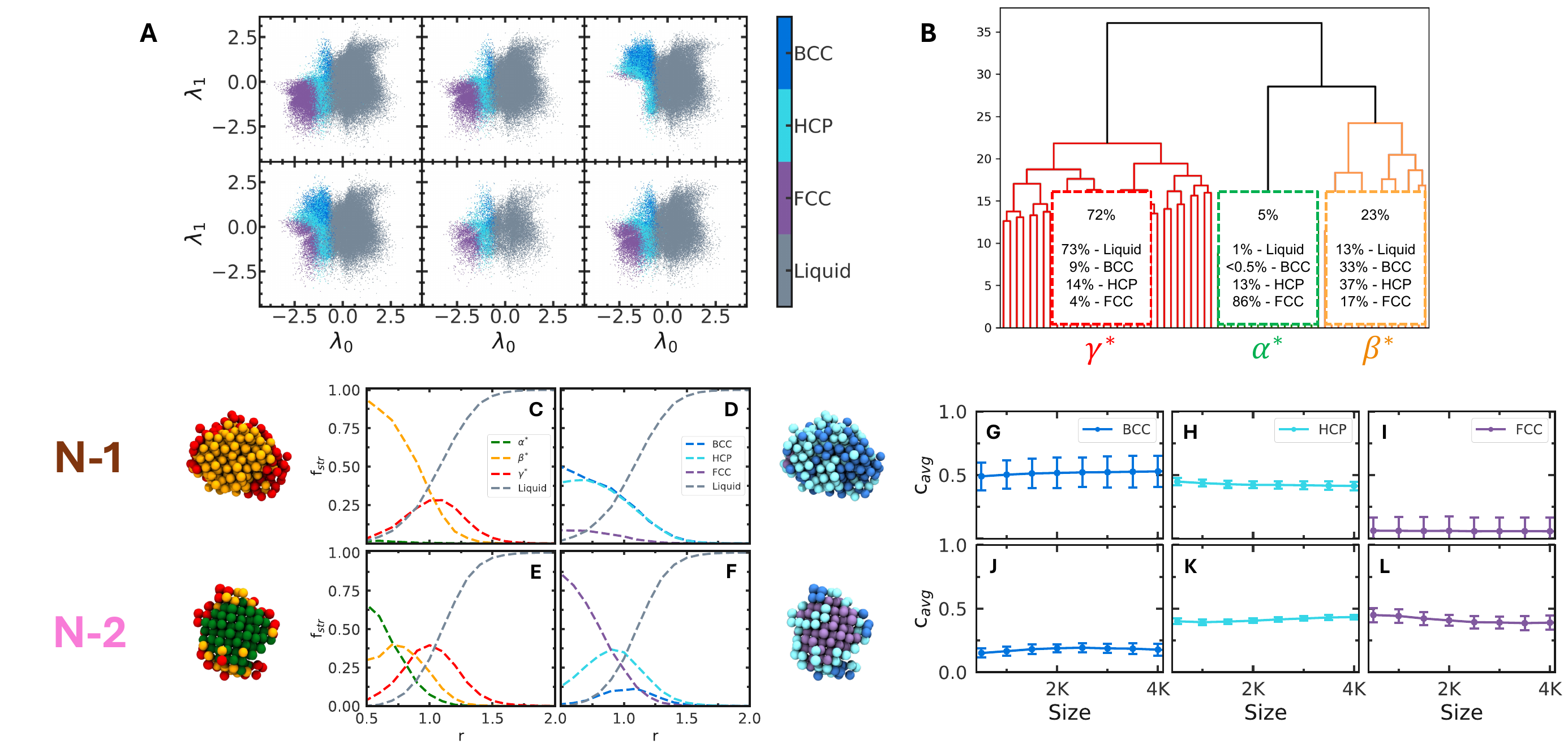}
\caption{\textit{(A)} Latent space projection of six Ni$_3$Al nucleation trajectories with phase labels. Each panel represents a 2D latent space projection of training data from each Ni$_3$Al nucleation trajectory with phase labels. Phase labels are based on $\overline{q}_4 $ and $ \overline{q}_6$ values. \textit{(B)} Particle path clusters for Ni$_{3}$Al system. Each colored region of the dendrogram represents a cluster of particle paths ($\alpha^{*}$, $\beta^{*}$, $\gamma^{*}$). \textit{(C-F)} Radial composition analysis of Ni$_3$Al configurations from each trajectory cluster in \textit{SI Appendix}, Fig.~S10. RCA of panel C and E use particle path cluster labels from panel B ($\alpha^{*}$, $\beta^{*}$, $\gamma^{*}$), whereas panel D and F use labels according to $\overline{q}_4 $ and $ \overline{q}_6$ values. The radial compositions are computed for (C-D) N-1 cluster, and (E-F) N-2 cluster from \textit{SI Appendix}, Fig.~S10. Configurations with committor estimates greater than or equal to 0.9 are selected for the calculation. \textit{(G-L)} Average composition of Ni$_3$Al nucleation trajectories from each trajectory cluster in \textit{SI Appendix}, Fig.~S10. Blue -- BCC composition, cyan -- HCP composition, purple -- FCC composition. The composition is calculated for (G-I) trajectories in the N-1 cluster and (J-L) trajectories in the N-2 cluster. The error bars reflect the standard deviation of the average composition across all trajectories launched within each trajectory cluster.}
\label{fig:ni3al-overall-results}
\end{figure*}

To evaluate whether LeaPP can learn the local atomic environments of Ni$_{3}$Al nucleation particles, we examine the placement of the local structures in the latent space after training. Figure \ref{fig:ni3al-overall-results}A illustrates the projection of training samples from six Ni$_{3}$Al nucleation trajectories onto the latent space using the trained VAE model. The labels of each point within the latent space are calculated from $\overline{q}_4$ and $\overline{q}_6$. Much like the 7--6 system, the VAE learns the local environments of different atom types and distinctly places them in the latent space. Liquid atoms are distinguished from solid atoms, and solid atoms are individually delineated. FCC atoms are the furthest from liquid atoms, and are most separated from BCC and HCP solids. In certain sample trajectories, BCC atoms exhibit overlaps with HCP atoms within the latent space. Overall, the relative distribution of different atom types within each trajectory's latent space projection demonstrates LeaPP's ability to capture environmental differences of particles during nucleation.  

\subsection*{Ni$_3$Al particle path clustering}

Similar to clustering particle paths in the 7--6 system, we project particle paths onto the latent space and cluster them based on similarities in their time evolution.  Particles in their last frame are labeled with $\overline{q}_4 $ and $ \overline{q}_6$, as explained for the 7--6 system. The distance cutoff for clustering is chosen such that there are three distinct particle path clusters, which we label as $\alpha^{*}$, $\beta^{*}$ and $\gamma^{*}$ (Fig. \ref{fig:ni3al-overall-results}B). The $\gamma^{*}$ cluster, which accounts for the highest number of particles, exhibits a high percentage of liquid. Notably, $\alpha^{*}$ and $\beta^{*}$ particle path clusters, representing about a third of the tracked particles, exhibit a higher percentage of solids in their definition at the last frame. The $\beta^{*}$ cluster, the second largest, is mainly composed of BCC and HCP atoms. The $\alpha^{*}$ cluster, with the fewest particles, predominantly consists of FCC particles. 

Next, we perform RCA on configurations from all Ni$_3$Al nucleation trajectories to understand the behavior of the particles based on their spatial correlation with the solid nucleus. RCA results show that $\alpha^{*}$ and $\beta^{*}$ particles correspond to atoms at the core of the solid nucleus, while the $\gamma^{*}$ particles make up the interfacial region of the nucleus (\textit{SI Appendix}, Fig.~S9). In the $\overline{q}_4 $ and $ \overline{q}_6$ interpretation of these different layers of the nucleus, the RCA shows that the interfacial structure mainly exhibits a mixture of BCC and HCP particles. As for the core particles, which consist of $\alpha^{*}$ and $\beta^{*}$ particles, together they make up a mixture of the three solid phases. 
\begin{figure}
\centering
\includegraphics[width=\linewidth]{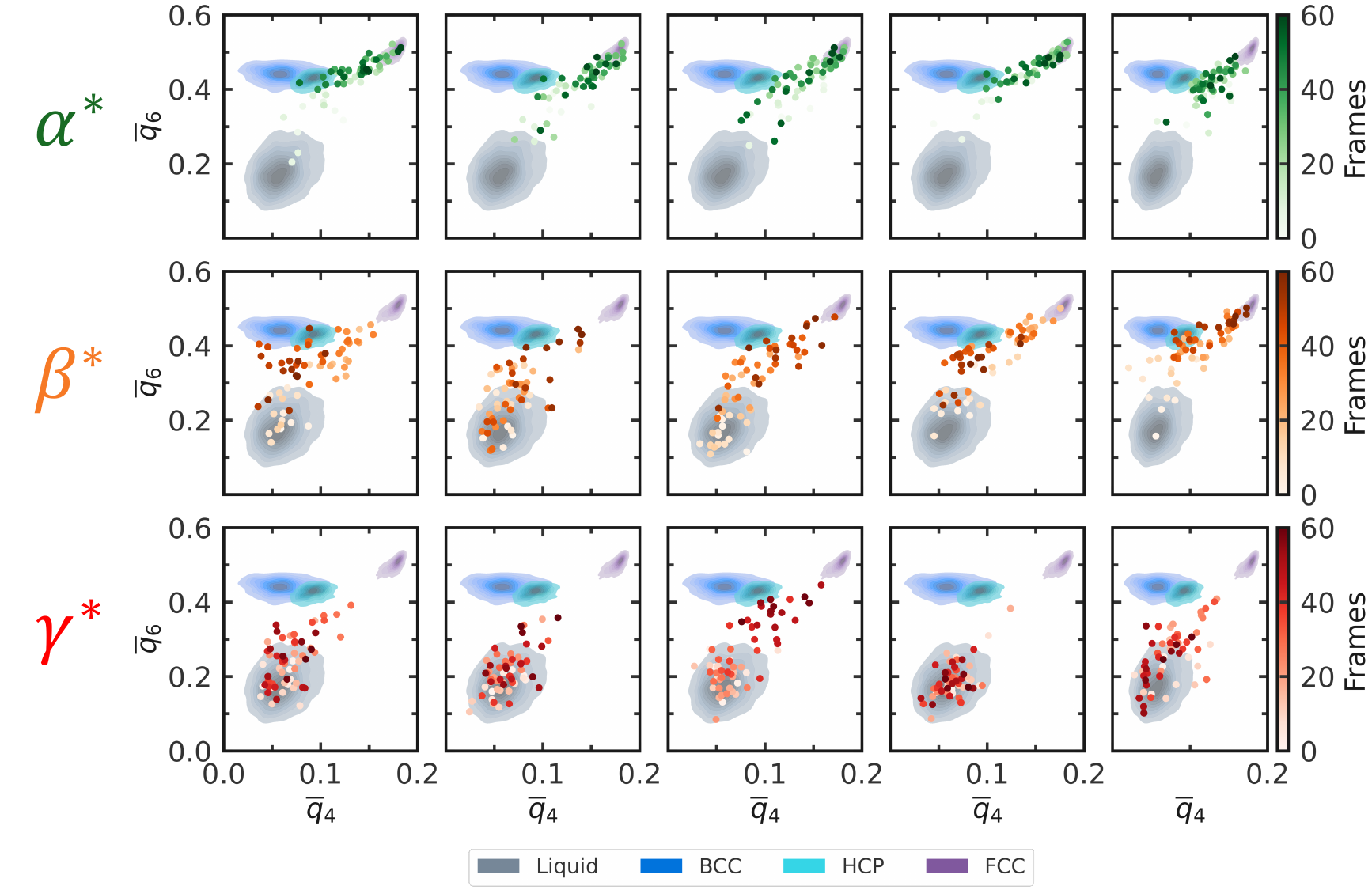}
\caption{Ni$_3$Al particles time evolution in $\overline{q}_{4}$ and $\overline{q}_{6}$ space. Each row represents time evolution plots for each particle path type: $\alpha^{*}$ (green), $\beta^{*}$ (orange) and $\gamma^{*}$ (red). Each plot demonstrates the time evolution of a particle with its associated cluster identity. The color bar reflects the time evolution in frames.}
\label{fig:pp-time-evo}
\end{figure}
The latent space separation between liquid and solid particles and our analysis of the time evolution of particles allows LeaPP to learn the difference between interfacial and core particles of the nucleus. We verify findings from LeaPP by tracing these particles time evolutions in the $\overline{q}_{4}$ and $\overline{q}_{6}$ space. We observe that the time evolution of $\gamma$ particles is transient and shows a fluctuation in and out of the liquid region (Fig. \ref{fig:pp-time-evo}). These fluctuations from liquid to solid allow LeaPP to differentiate these $\gamma$ particles from the $\alpha^{*}$ and $\beta^{*}$ particles, which emerge as core particles of the solid nucleus. Unlike $\gamma$ particles, they do not show fluctuations from liquid to solid and vice versa. These particles remain as solid atoms once they transition from liquid. Both the RCA results and the time evolution plots demonstrate the utility of considering the time evolution of particles in characterizing their identity. 

\subsection*{Ni$_3$Al nucleation trajectory clustering}

Using the distinctive signatures of the particle path clusters, our subsequent step involves analyzing the relative abundance of $\alpha^{*}$, $\beta^{*}$, and $\gamma^{*}$ in each trajectory to effectively cluster the Ni$_{3}$Al nucleation trajectories. From simulation trajectories extending from the critical size in Ref.~\cite{Liang2020}, Liang et al. discovered two polymorph profiles from {Ni$_3$Al} nucleation trajectories. Utilizing the cutoff of 0.8, LeaPP also clusters the nucleation trajectories into two clusters: N-1 and N-2 (see \textit{SI Appendix}, Fig.~S10). RCA results show that configurations from N-1 cluster have a dominant $\beta^{*}$ core and $\gamma^{*}$ interface (Fig. \ref{fig:ni3al-overall-results}C), while configurations from N-2 cluster have a dominant $\alpha^{*}$ core with $\beta^{*}$ particles layered in between the $\alpha^{*}$ core and $\gamma^{*}$ interfacial particles (Fig. \ref{fig:ni3al-overall-results}E). In the $\overline{q}_{4}$ and $\overline{q}_{6}$ interpretation, configurations of the N-1 trajectory cluster have a mixture of BCC and HCP throughout the nucleus, with traces of FCC at the core. In contrast, configurations of N-2 trajectory cluster exhibit a dominant FCC core and HCP interface mixed with few BCC atoms. The resulting two radial composition profiles demonstrate the possibility of two nucleation pathways. Thus, we extend five trajectories from the final configuration of each nucleation trajectory. Fig. \ref{fig:ni3al-overall-results}G-L demonstrates that LeaPP successfully identifies two distinct polymorphic profiles. Notably, while the HCP content remains relatively consistent between N-1 and N-2 trajectories, the reciprocal relationship between BCC and FCC content within the nucleus governs the predominant polymorph during the growth process. Specifically, N-1 trajectories exhibit a higher proportion of BCC compared to FCC, while N-2 trajectories demonstrate the opposite trend. In line with the results from Liang et al. \cite{Liang2020}, our method detects two distinct nucleation pathways in an unsupervised manner. One pathway is characterized by the dominance in BCC-HCP content and the other in FCC-HCP content. 

However, as demonstrated for the 7--6 potential system, further decreasing the cutoff can expand the discovery of nucleation trajectory clusters. Upon further subdivision of the N-2 cluster, two trajectory clusters emerge: N-2.1 and N-2.2 trajectory cluster (\textit{SI Appendix}, Fig.~S11). However, their growth evolution in \textit{SI Appendix}, Fig.~S13D-F (N-2.1 cluster) and S13G-I (N-2.2 cluster) demonstrate a consistent resemblance in their polymorph profiles based on the $\overline{q}_4 $ and $ \overline{q}_6$ interpretation. 

\section*{Discussion} 
In nucleation, crystal nuclei exhibit a wide diversity of structures as they form and grow from the liquid. By tracing the time evolution of the local structures around particles, we demonstrate the ability of LeaPP to characterize nucleation trajectories and their constituent particles in an unsupervised manner. Through this method, we encapsulate the relationship between the structure and dynamics of nucleating particles to unravel their roles in nucleation. On the one hand, LeaPP can differentiate between interfacial and core particles of the nuclei. This capability to consistently separate surface particles from the core is exciting as it allows us to study the properties of these interfacial particles without relying on traditional order parameters that are based on bulk structure definitions. Instead, we can analyze interfacial particles by correlating them with other structural and dynamical properties, gaining insights into their role in nucleation growth or melt. For the systems tested above, we observe that particles that join the core of the nucleus evolve from liquid to solid and remain fluctuating in the solid region. Whereas, the interfacial particles fluctuate in and out of the liquid phase, demonstrating their transient role in nucleation.

Additionally, for the systems tested here, LeaPP also identifies nucleating particles that proceed through different local structures. LeaPP adeptly highlights subtle differences between particle paths and accurately categorizes nucleation trajectories in an unsupervised approach. Using a conservative cutoff in the 7--6 potential system, we uncover two distinct pathways to different polymorphs: "BCC-dominant" trajectories and "FCC-HCP-dominant" trajectories. LeaPP's precision in predicting growth phase polymorph selection surpasses that of a traditional order parameter. The findings from LeaPP indicate a wide range of possible structures that nuclei can sample. This is consistent with the results of similar systems such as LJ particles \cite{Bolhuis:05:PRL}, inverse power law potentials \cite{Delhommelle2007JPCB}, and the Gaussian core model \cite{Bolhuis2011JCP,Tanaka2012SM,Sear2015JCP}, where BCC, HCP, and FCC structures all contribute to the ensemble of nucleation pathways. In the Ni$_{3}$Al system, we confirm that the reciprocal relationship between the BCC and FCC phase dictates the selection of polymorph during growth. Two distinct structural pathways (BCC-HCP dominant and FCC-HCP dominant) are found and this result agrees with the previous findings of Liang et al. \cite{Liang2020}. Overall, while not every identified trajectory cluster represents a distinct polymorph at the crystalline stage upon relaxing the cutoff criteria, the differences in nucleus composition detected by LeaPP during growth offer a more detailed understanding of the diverse pathways a growing nucleus could take.

While we chose to use Steinhardt order parameters in this work, the choice of local descriptors to use as input is general. Machine learning methods have recently enabled more precise characterization of the local environments of molecular simulations. Any features learned from the local environments, such as through unsupervised methods \cite{FilionJCP2019,Glotzer:20:JPCB,YasuokaJCTC2024,GrossmanNC2019}, can in principle be used in conjunction with LeaPP. Becchi et al. \cite{Pavan2024arXiv} recently developed a method to identify and cluster the infrequently sampled local environments. By combining the strengths of such methods with LeaPP, we envision an approach to characterize the pathways for a wide array of self-assembly problems.

\section*{Materials and Methods}
\subsection*{Input features} \label{input-features}

Steinhardt BOPs \cite{SteinhardtPRB1983}, $q_l$, and their neighbor-averaged variants, $\overline{q}_l$ \cite{Dellago:08:JCP}, are calculated to characterize the local structure around each atom. These order parameters are commonly used to distinguish between structures in colloidal systems and water. They also possess invariance to the rotation, translation, and permutation of the local atomic environment. $q_l$ and $\overline{q}_l$ are calculated for $l=\{2,4,6,8\}$ and up to four cutoff distances $r_\text{cut}$. For a given value of $l$ and $r_\text{cut}$, each of the $2l+1$ components of the vector $\mathbf{q}_l$ is calculated for each atom $i$ based on the spherical harmonics:
\begin{equation}
    q_{lm}(i) = \frac{1}{N_\text{neigh}(i)} \sum_{j=0}^{N_\text{neigh}(i)}{Y_{lm}(\mathbf{x}_j)},
\end{equation}
where $N_\text{neigh}$ is the number of neighboring atoms within $r_\text{cut}$ of atom $i$, and $Y_{lm}$ are the spherical harmonics functions. $\mathbf{x}_j$ is the relative position of neighbor $j$ to atom $i$. To calculate the neighbor-averaged $q_l$, neighbor averaging is first applied with
\begin{equation}
    \overline{q}_{lm}(i) = \frac{1}{N_\text{neigh}(i)+1} \sum_{j=0}^{N_\text{neigh}+1}{q_{lm}(j)},
\end{equation}
where the additional term in the sum is $q_{lm}(i)$ for the central atom. Finally, $q_l$ is calculated as
\begin{equation}
    q_l(i) = \sqrt{\frac{4\pi}{2l+1} \sum_{m=-l}^{l}{|q_{lm}(i)|^2}}
\end{equation}
Similarly, the neighbor-averaged equivalent is calculated as
\begin{equation}
    \overline{q}_l(i) = \sqrt{\frac{4\pi}{2l+1} \sum_{m=-l}^{l}{|\overline{q}_{lm}(i)|^2}}
\end{equation}

\subsection*{Training data}

Our method is applied to two systems---the 7--6 potential and Ni$_3$Al. To sample a broad range of local environments that may be present during nucleation, replica exchange transition interface sampling (RETIS) is applied, resulting in 181 and 82 nucleation trajectories for the 7--6 and Ni$_3$Al systems, respectively. In all frames of the nucleation trajectories, the maximum size of the largest solid nucleus is typically less than 10\% of the total atoms in the system. Therefore, we balance the sampling towards solid atoms by only considering atoms that are part of the largest solid nucleus for at least one frame of the trajectory. The largest solid nucleus in every frame is calculated using $\overline{q}_6$ for each nucleation trajectory. For the 7--6 potential, neighbors are defined with $r_\text{cut}=1.28 \sigma$, where $\sigma$ is the distance unit, and the cutoff for $\overline{q}_6$ to define solid atoms was 0.36. The same $r_\text{cut}$ is used to cluster the solid atoms and calculate the largest nucleus size. If an atom is ever part of the largest solid nucleus for at least one frame, its local environment is used to calculate the BOP features for the entire trajectory.

For the atoms that are tracked, BOPs are calculated every 0.4 $\tau$ for 7--6 particles. Three values of $r_\text{cut}$ were selected to calculate the BOPs. These were selected to account for (i) the first neighbor shell of the bulk FCC phase, (ii) the first neighbor shell of the liquid phase, and (iii), the first two neighbor shells of the bulk FCC phase. These correspond to $r_\text{cut}$ values of 1.25, 1.40 and 1.60 $\sigma$. Once the BOPs are calculated, there is a total of 24 BOPs per 7--6 particle. BOPs for the Ni$_3$Al atoms were calculated every 1 ps, using $r_\text{cut}$ values of 0.30, 0.35, 0.47 and 0.55 nm.

\subsection*{Dimensionality reduction}

A variational autoencoder (VAE) is used to obtain a low-dimensional projection of the local atomic environments. The models in this work use fully-connected layers for the encoder and decoder. The encoder consists of two hidden layers that reduce the input dimensionality to that of the latent space, $d$, where $\mu$ and $\log(\sigma^2)$ are outputs, which are the means and log variances. The decoder also consists of two hidden layers to reconstruct the input features. The encoder hidden layers contain 32 and 16 nodes, while the decoder hidden layers contain 16 and 32 nodes. For 7--6, $d=3$, and $d=2$ for Ni$_3$Al. Batch normalization and ReLU are applied to each layer, with the exception of the layers that output the latent space values and the reconstructions. Additional information on the VAE model is available in the \textit{SI Appendix}.

Prior to training the VAE on the 7--6 system, the input data were balanced based on the values of $\overline{q}_4$ and $\overline{q}_6$ such that each bulk phase was roughly equally represented. For training, 200 particles were chosen from each of the liquid, BCC, HCP and FCC phases from each nucleation trajectory. For the Ni$_3$Al system, the full set of atoms from all nucleation trajectories was used and split between training (80\%) and validation (20\%) data.

\subsection*{Nucleation trajectory characterization}

Once the VAE is trained, the atomic feature dataset is organized into particle paths, which is the time evolution of each atom's features. Each atom's local environment is projected into the VAE latent space at each frame.

The particle paths can have variable lengths, so dynamic time warping (DTW) \cite{ChibaITASS1978} is used to calculate distances between pairs of paths. DTW seeks to find a correspondence between frames of two time series with (possibly) variable length, and the DTAIDistance \cite{dtaidistance} package for Python is used here. DTW has been previously used to distinguish between trajectories of molecular configurations \cite{HummerNCS2023,ParrinelloARXIV2023}. Here, multi-dimensional DTW \cite{KeoghDMKD2017} is applied using the latent space coordinates. Distances were calculated for every pair of particle paths in the VAE latent space. This distance matrix is then used as the distance metric for hierarchical agglomerative clustering with complete linkage. The number of clusters for the particle paths is generally selected between three and five, depending on the structures of the resulting dendrograms. The resulting clusters then represent atoms based on how they evolve over the course of nucleation and the early stages of growth. For a comparison with more conventional structures, the particle path clusters are characterized by the correspondence of their constituent atoms with known bulk phases---here, liquid, FCC, HCP, and BCC.

Finally, nucleation trajectories are clustered based on the type of particles that exist within the trajectory. For each nucleation trajectory, a histogram of particle types from each frame of the trajectory is constructed. The difference between two trajectories' histograms is calculated as the "distance" between the two trajectories. The resulting distance matrix is used with hierarchical agglomerative clustering to generate groups of nucleation trajectories.

\subsection*{Simulation details}

The 7--6 potential is from the work of Ref.\cite{Shi_Structure_2011}, with the form
\begin{equation}
V(r)=\epsilon\Bigg[\theta\Big(\frac{\sigma}{r}\Big)^n-\psi\Big(\frac{\sigma}{r}\Big)^6\Bigg],
\end{equation}
where $n=7$, $\theta=13.468$ and $\psi=14.000$. The potential has the same general shape as the Lennard-Jones 12--6 (LJ) potential, but the repulsive and attractive forces near the minimum in the potential are lesser, giving a softer potential. The melting temperature at a pressure of 5.0 was found with the direct coexistence \cite{Fernndez_The_2006} method to be 1.07 (see Supporting Information for direct coexistence simulation details). Simulations were carried out with 8192 particles and at 78\% of the melting temperature, or $T=0.83$, using the constant particle number, pressure and temperature ensemble in GROMACS 5.1.2 \cite{Abraham_GROMACS_2015}. Simulation settings were otherwise identical to the LJ simulations with replica exchange transition interface sampling (RETIS) in Ref. \cite{SarupriaJCP2022}.

Nucleation trajectories are sampled with RETIS \cite{Bolhuis:08:JCP1,vanErp:07:PRL}. This ensures that the simulations are not influenced by a bias potential and allows us to sample a large number of decorrelated trajectories. For RETIS with the 7--6 potential, the transition region is divided into 17 interface ensembles by positioning RETIS interfaces at $\lambda$ = \{45, 55, 70, 85, 100, 115, 130, 145, 170, 190, 220, 250, 280, 320, 360, 410, 520\}, and the final interface is placed at $\lambda = 720$. $\lambda$ corresponds to the size of the largest nucleus based on $q_6$ from Ref. \cite{ten_Wolde_Numerical_1996, tenWolde:05:PRL} (see Supporting Information for calculation details). After an initial trajectory is generated for each ensemble, RETIS attempts to either exchange trajectories between neighboring interfaces or generate a new trajectory with a shooting move. Here, we use 50\% shooting moves and 50\% exchange moves. For the shooting move, velocities are drawn from the Maxwell--Boltzmann distribution and trajectories are propagated forward and backward from the shooting point configuration with two-way shooting. Within interface ensemble $i$, the shooting point configuration is uniformly selected among configurations between $i-1$ and $i+1$. In total, we perform 5000 MC moves for each interface ensemble and discard the first 600 trajectories from our analysis. 

LeaPP is also done using Ni$_3$Al nucleation trajectories sampled with RETIS in Ref. \cite{Liang2020}. Simulations were run at 0 bar and 1342 K, corresponding to a 20\% supercooling at 0 bar. The system consisted of 6912 atoms with 75\% Ni and 25\% Al. Interactions used an embedded atom potential developed for Ni--Al \cite{Mishin:09:PhilMag}.

\subsection*{Committor estimation from RETIS}

From RETIS, estimates of the committor can be obtained by projecting the reweighted path ensemble onto a set of order parameters using the scheme from Rogal et al. \cite{Bolhuis:10:JCP}. The chosen order parameters should closely correspond to the reaction coordinate for a more accurate projection. From maximum likelihood estimation \cite{TroutJCP2006, PetersBook:17}, the combination of two order parameters: the largest nucleus size and the average crystallinity of the nucleus (based on $q_{6}$) \cite{ten_Wolde_Numerical_1996, tenWolde:05:PRL} were found to give the best two-component reaction coordinate (the procedure for calculating the largest nucleus size and the degree of crystallinity of a nucleus is described in the methods section of the Supporting Information). The committor projection of the reweighted path ensemble is then projected along these two order parameters in bins. A configuration is assigned the committor of the bin it belongs to.

\section*{Acknowledgements}
Jutta Rogal and Yanyan Liang are acknowledged for sharing their simulation data and useful discussions. S.W.H. and S.S. acknowledge the Data Science Initiative at the University of Minnesota College of Science and Engineering for funding through the ADC Graduate Fellowship. This material is based on work supported by NSF CBET No. 1653352. This work was supported by the U.S. Department of Energy, Office of Science, Office of Basic Energy Sciences, under Award No. DE-SC0015448. The authors acknowledge the University of Minnesota Start-up funds and the Minnesota Supercomputing Institute at the University of Minnesota for providing resources that contributed to the reported results.

\bibliographystyle{tfnlm}
\bibliography{arxiv-main.bbl}

\end{document}


\maketitle

\section*{Direct coexistence simulations}
\begin{figure}
\centering
\includegraphics[width=0.7\textwidth]{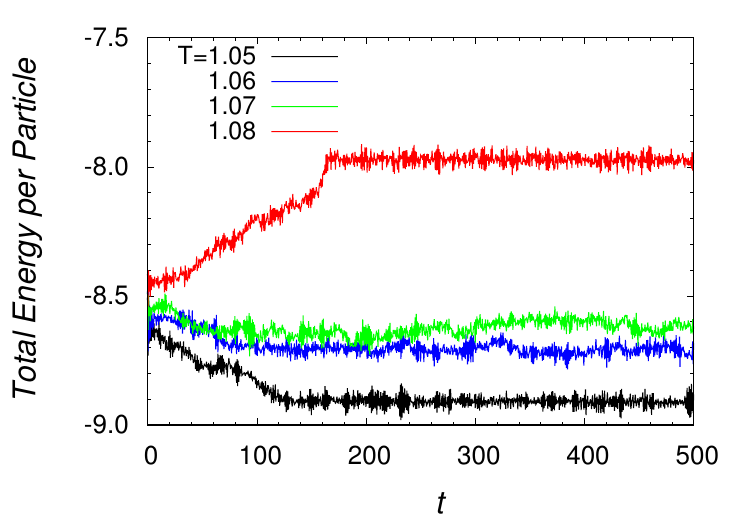}
\caption{Evolution of total energy per particle in the 7-6 direct coexistence simulations at a pressure of 5.}
\label{fig:76dc-sim}
\end{figure}

At each pressure of interest, direct coexistence simulations \cite{Vega2006JCP} were run across a range of temperatures. The initial configuration was generated by contacting a liquid configuration of 4000 particles with the (100) plane of an equilibrated FCC crystal containing 3872 particles. Energy minimization was performed on the liquid particles to remove high-energy overlaps at the interfaces, followed by a 500-$\tau$ $Np_zT$ simulation where box fluctuations were only allowed in the z-direction. At temperatures above the melting temperature, the solid region melts and the total energy increases as the thermostat resupplies kinetic energy to the system. The opposite occurs below the melting temperature, causing a decrease in the energy. By exploiting this behavior, the melting temperature is determined by finding the lowest temperature at which the solid melts and the highest at which it grows. The two are averaged to obtain the melting temperature with an error equal to half the difference between the temperatures. Example energy plots from simulations are shown in Fig. \ref{fig:76dc-sim}.

\section*{The size of the largest nucleus ($n_{\text{tf}}$)}

The overall procedure for calculating the largest nucleus size is as follows: identify solid particles from the degree of structural correlation with neighboring particles, and then identify the largest cluster of solid particles. First, the neighbors of each particle are identified. Two particles $i$ and $j$ are considered neighbors if their distance is less than some cut-off distance, $|\textbf{r}_{ij}| < r_\text{cut}$, where $r_\text{cut} = 1.42\sigma$. The vertical bars indicate an $l^2$ norm. Solid particles are identified using a procedure similar to that of ten Wolde \textit{et al.} \cite{ten_Wolde_Numerical_1996}.

A complex vector, $q_{6m}$ of each particle $i$ is calculated as
\begin{equation}
q_{6m}(i)=\frac{1}{N_n(i)}\sum_{j=1}^{N_n(i)} Y_{6m}(\textbf{r}_{ij}),
\label{eq:q6m}
\end{equation}
where $Y_{6m}$ are the $l=6$ spherical harmonics, $\textbf{r}_{ij}$ is the unit vector between particles $i$ and $j$, and the sum is over the $N_n$ neighbors of particle $i$.  The resulting complex vector is 13-dimensional since the $l=6$ spherical harmonics span integers from $m=-6$ to $m=6$. The normalized $q_{6m}$ dot product of each neighbor pair,
\begin{equation}
d_{ij}=\frac{\sum_{m=-6}^{m=6} q_{6m}(i)q_{6m}(j)^{*}}{\big(\sum_{m=-6}^{m=6} |q_{6m}(i)|^2\big)^{1/2}\big(\sum_{m=-6}^{m=6} |q_{6m}(j)|^2\big)^{1/2}}
\end{equation}
is used to determine the degree of structural correlation between particles. The asterisk denotes a complex conjugate. Two particles $i$ and $j$ are considered connected if $d_{ij} > d_\text{cut}$, where $d_\text{cut} = 0.5$. The total number of connections particle $i$ has with neighboring particles is given by
\begin{equation}
n_\text{connec}(i) = \sum_{j=1}^{N_n(i)} h(d_{ij} - d_\text{cut})
\end{equation}
where $h(\cdot)$ is the Heaviside step function. Particle $i$ is considered solid if $n_\text{connec}(i) \ge n_\text{connec,cut}$, where $n_\text{connec,cut} = 9$.

Once all solid particles in the system are identified with the above procedure, the largest cluster of solid particles is identified. A graph is constructed with solid particles as nodes. Edges are placed between nodes which are neighbors (i.e., $|\textbf{r}_{ij}| < r_\text{cut}$). The largest cluster of solid particles is the largest connected component of the graph. $n_\text{tf}$ is the number of particles in the largest cluster.

\section*{Degree of crystallinity of the nucleus ($Q_6^{\text{cl}}$)}
The measure of the degree of crystallinity of a nucleus is calculated by averaging the $Y_{6m}$ spherical harmonics over all bonds between solid neighbors in the cluster \cite{Moroni_Interplay_2005}. This gives the 13-dimensional complex vector $Q_{6m}^{\text{cl}}$. $Q_6^{\text{cl}}$ is then calculated as
\begin{equation}
Q_6^{\text{cl}}(i)=\bigg(\frac{4\pi}{13}\sum_{m=-6}^{m=6} |Q_{6m}^{\text{cl}}(i)|^{2}\bigg)^{1/2}.
\end{equation}

\section*{Variational autoencoder architecture (VAE)}
For dimensionality reduction, a $\beta$-VAE \cite{HigginsICLR2017} model was used to modulate the relative weights between the reconstruction loss and latent space regularization. A value of $\beta=0.06$ was chosen for our models. Both $\beta$ and the latent space dimensionalities were chosen based on visualization of the latent space. The architectures for the VAE models are shown in Tab.~\ref{tab:vae}.
\begin{table}[h]
\linespread{0.9}\selectfont\centering
\caption{VAE architectures for the 7--6 and Ni$_3$Al systems. Numbers are the number of components in the output of a particular layer. In the latent space, the mean and log variances are generated from two different linear layers.}
\label{tab:vae}
\begin{center}
\setlength{\tabcolsep}{9pt}

\begin{tabular}{@{} l c c @{}} \toprule

& \multicolumn{2}{c}{System}\\
\cmidrule{2-3}
Layer & 7--6 & Ni$_3$Al  \\
\midrule
Input & 24 & 32 \\
Linear1 & 32 & 32 \\
Batchnorm1 & 32 & 32 \\
Relu1 & 32 & 32 \\
Linear2 & 16 & 16 \\
Batchnorm2 & 16 & 16 \\
Relu2 & 16 & 16 \\
Linear3 (Latent space) & 2 $\times$ 3 & 2 $\times$ 2 \\
Linear4 & 16 & 16 \\
Batchnorm4 & 16 & 16 \\
Relu4 & 16 & 16 \\
Linear5 & 32 & 32 \\
Batchnorm5 & 32 & 32 \\
Relu5 & 32 & 32 \\
Output & 24 & 32 \\
\bottomrule

\end{tabular} 
\end{center}
\end{table}

\subsection*{Radial composition analysis (RCA) calculation}
In RCA, the fraction of each particle type is calculated for each shell of the nucleus, starting from the center of the nucleus until bulk liquid. The equation below describes the calculation of the fraction of particle type ($f_{P}$) in a shell $dr$ of the nucleus across $M$ number of configurations:
\begin{equation}
    f_{P, dr} = \frac{\frac{w_{i}\sum_{i}^{M} n_{P,dr,i}}{N_{\text{tot},dr,i}}}{\sum_{i}^{M} w_{i}}
\end{equation}
where $n_{P, dr, i}$ is the number of particles type $P$ within the shell $dr$ of configuration $i$, $N_{\text{tot}, dr, i}$ is the total number of particles within the shell $dr$ of configuration $i$  and $w_i$ is the weight of configuration $i$ from the reweighted path ensemble\cite{Bolhuis:10:JCP}.

In this work, we choose one configuration from each nucleation trajectory to perform the analysis. This is to ensure that there is a balanced contribution from each nucleation trajectory to the result, independent of the nucleation trajectory length. The idea is to not have the result be influenced by a subset of trajectories, which have more $p_{B} \geq 0.9$ configurations due to its length.

\subsection*{Maximum Likelihood Estimation}
We find the optimal reaction coordinate using the protocol detailed in Ref. \cite{TroutJCP2006,Bolhuis2011JCP,Bolhuis:10:JCP}. The choice of the model committor is 
\begin{equation}
    p_{B} = \frac{1+\tanh[r(\textbf{q})]}{2}
\end{equation} 
and the choices of the trial reaction coordinate ($r(\textbf{q})$) are:
\begin{equation}
    r(\textbf{q}) = \alpha_{1}q_{1}+\alpha_{0}
    \label{eq:trc-1}
\end{equation}
\begin{equation}
   r(\textbf{q}) = \alpha_{2}q_{2}+\alpha_{1}q_{1}+\alpha_{0}
   \label{eq:trc-2}
\end{equation}
\begin{equation}
    r(\textbf{q}) = \alpha_{1}q_{1}q_{2}+\alpha_{0} 
    \label{eq:trc-3}
\end{equation}
where $\alpha$ is a parameter and $q$ is an order parameter listed in Table \ref{tab:op-list}. Equation \ref{eq:trc-3} is found to be the optimal form of the reaction coordinate with $n_\text{tf}$ and $Q_{6, \text{tf}}^{\text{cl}}$ as the order parameters. Table \ref{tab:mle-op} shows the top 3 candidates from each form of the reaction coordinate and their negative log-likelihood score.

\begin{table}[h]
\linespread{0.9}\selectfont\centering
\caption{List of order parameters used in MLE.}
\label{tab:op-list}
\begin{center}
\setlength{\tabcolsep}{9pt}

\begin{tabular}{@{} l @{}} \toprule

Order Parameters ($q$)\\
\midrule
$n_\text{tf}$ \\  
$n_\text{ld}$ \\
$f_\text{BCC}$ \\ 
$f_\text{HCP}$ \\ 
$f_\text{FCC}$ \\ 
$Q_\text{6, tf}^{\text{cl}}$ \\ 
$Q_\text{6, ld}^{\text{cl}}$ \\  
$\kappa_\text{tf}$ \\ 
$\kappa_\text{ld}$ \\ 

\bottomrule

\end{tabular} 
\end{center}
\end{table}

\begin{table}[h]
\caption{Top three ranked OPs for each choice of the trial RC ($r(\textbf{q})$)}
\centering
    
\begin{tabular}{@{} l c c c c c c @{}} \toprule

& \multicolumn{2}{c}{$r(\textbf{q})=\alpha_{1}q_{1}+\alpha_{0}$} & \multicolumn{2}{c}{$r(\textbf{q})=\alpha_{2}q_{2}+\alpha_{1}q_{1}+\alpha_{0}$} & \multicolumn{2}{c}{$r(\textbf{q})=\alpha_{1}q_{1}q_{2}+\alpha_{0}$} \\
\cmidrule{2-3}
\cmidrule{4-5}
\cmidrule{6-7}
Rank & OP & -LL score  & OP & -LL score & OP & -LL score\\

\midrule
1 & $n_\text{tf}$&$7.050\times10^{-5}$&  $n_\text{tf}$ + $n_\text{ld}$&$5.740\times10^{-5}$& $n_\text{tf}Q_\text{6, tf}^\text{cl}$&$1.230 \times 10^{-3}$\\
2 & $n_\text{ld}$&$7.780\times10^{-5}$&  $n_\text{tf}$ + $\kappa_\text{tf}$&$5.780\times10^{-5}$& $n_\text{ld}Q_\text{6, ld}^\text{cl}$&$1.298\times10^{-3}$\\
3 & $Q_\text{6, tf}^\text{cl}$&$2.110\times10^{-4}$&  $n_\text{tf}$ + $f_\text{FCC}$&$5.840\times10^{-5}$& $n_\text{ld}Q_\text{6, tf}^\text{cl}$&$1.337\times10^{-3}$\\
\bottomrule

\end{tabular} 
\label{tab:mle-op}
\end{table}

\bibliographystyle{tfnlm}
\bibliography{arxiv-si.bbl}
\newpage

\begin{figure}
    \centering
    \includegraphics[width=\textwidth]{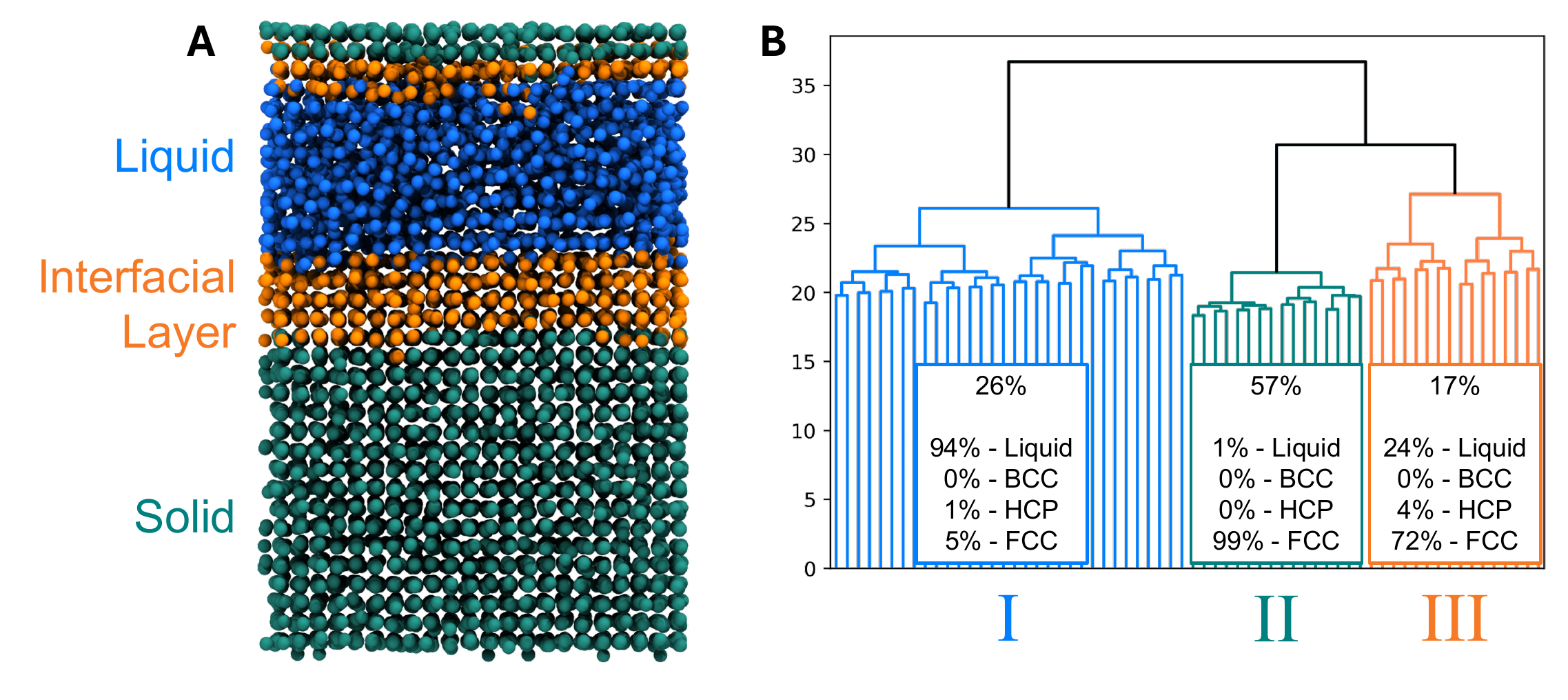}
    \caption{\textit{(A)} Snapshot of the last configuration from 7--6 liquid-solid direct coexistence trajectory. Blue particles correspond to liquid particles, orange particles correspond to particles in the interfacial layer, and forest green particles correspond to solid particles. \textit{(B)} Particle path clusters for 7--6 particles. Each colored region of the dendrogram represents a cluster of particle paths (I, II, III). The particle path similarities are calculated via DTW. DTW is used to cluster the particle paths via agglomerative hierarchical clustering. Particle path clusters that share the most similarity will join branches earlier as the cutoff (y-axis) is increased. Each dendrogram cluster’s composition is calculated from the final frame of all its constituent particle paths. The percentage at the top center of each cluster is the fraction of particle paths that belong to each cluster. The percentage of liquid, BCC, FCC, and HCP is calculated from the label of the particle in the last frame.}
    \label{fig:76coex}
\end{figure}

\begin{figure}
    \centering
    \includegraphics[width=0.5\textwidth]{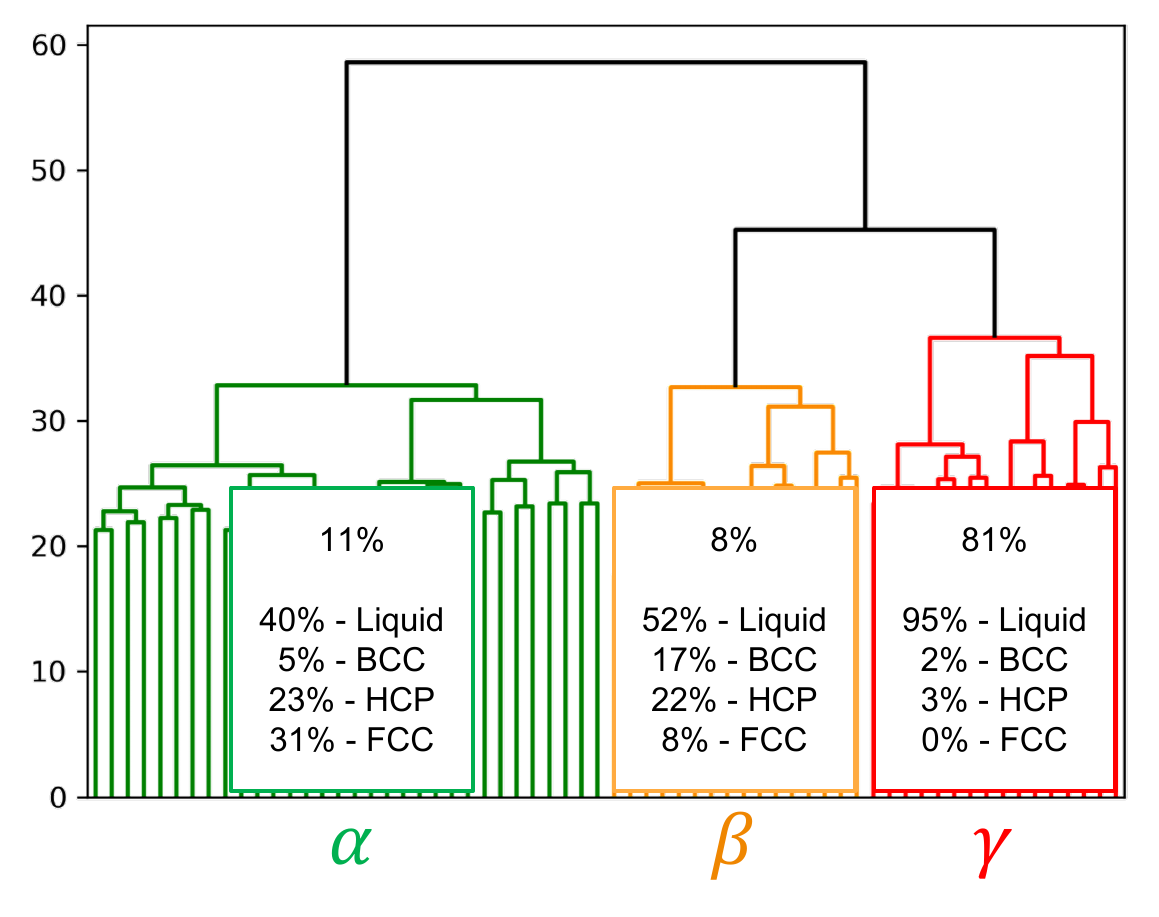}
    \caption{\textbf{Particle path clusters for 7–6 particles.} Each colored region of the dendrogram represents a cluster of particle paths ($\alpha$, $\beta$, $\gamma$). The percentage at the top center of each cluster is the fraction of particle paths that belong to each cluster. The percentage of liquid, BCC, FCC and HCP is calculated from the label of the particles averaged over all frames.}
    \label{fig:lj-ppc3-usingallframes}
\end{figure}

\begin{figure}
    \centering
    \includegraphics[width=0.8\textwidth]{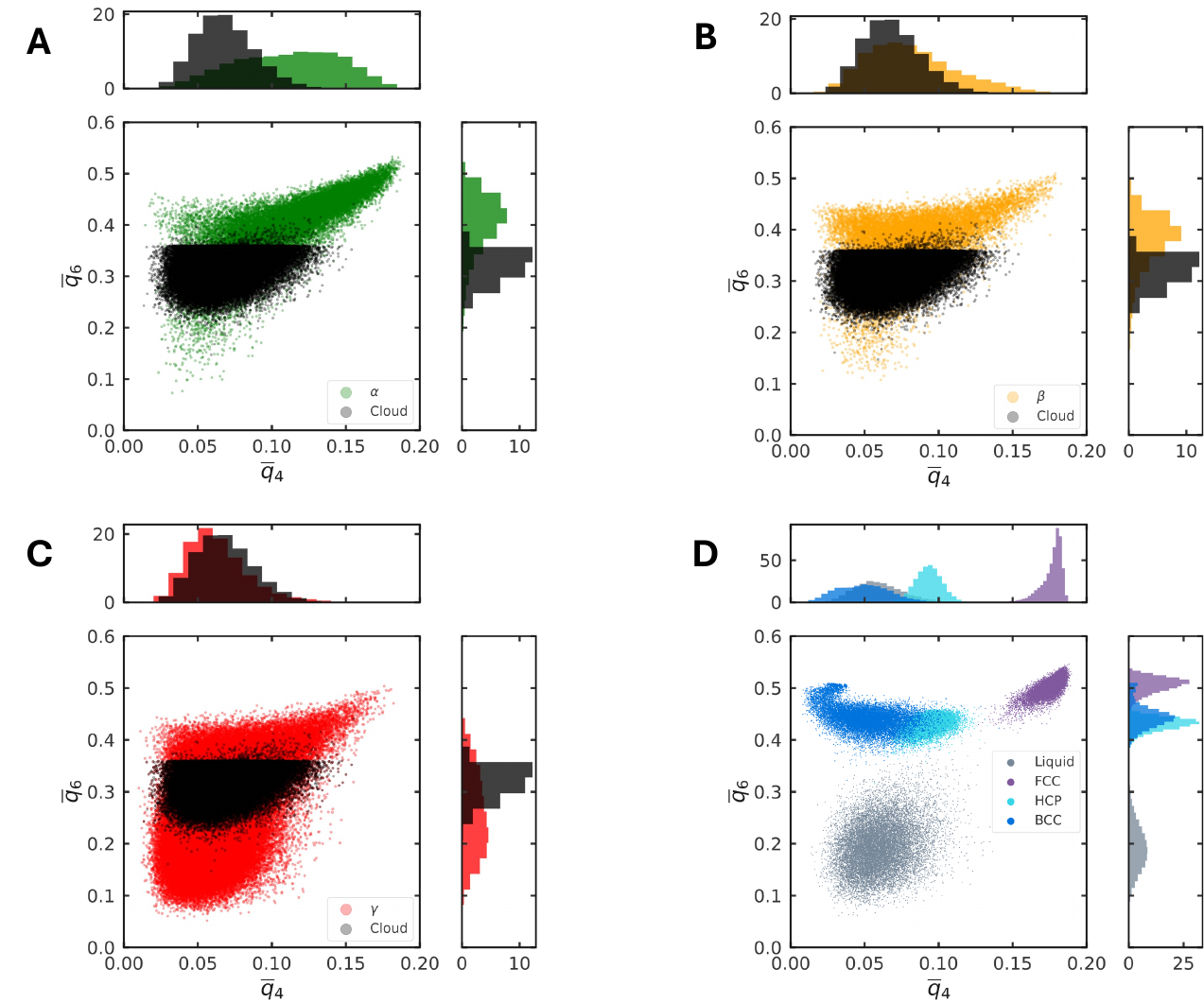}
    \caption{\textbf{7--6 particles projection in $\overline{q}_4$ and $\overline{q}_6$ space.} Panel A-C demonstrates the projection of "cloud particles" in black (as defined by Lechner et al. \cite{Lechner_Role_2011}) along with $\alpha$ particles (A), $\beta$ particles (B), and $\gamma$ particles (C) from Fig. 2. The particles are projected into the $\overline{q}_4$ and $\overline{q}_6$ space, where $r_\text{cut} = 1.28$. The $\alpha$, $\beta$, and $\gamma$ particles here include particles that are in the solid nucleus in the last frame as well as other tracked particles that are not a part of the solid nucleus. Panel D shows reference bulk structures in the $\overline{q}_4$ and $\overline{q}_6$ space with the same $r_\text{cut}$.}
    \label{fig:cloud-particles-correlation}
\end{figure}

\begin{figure}
\centering
\includegraphics[width=0.7\textwidth]{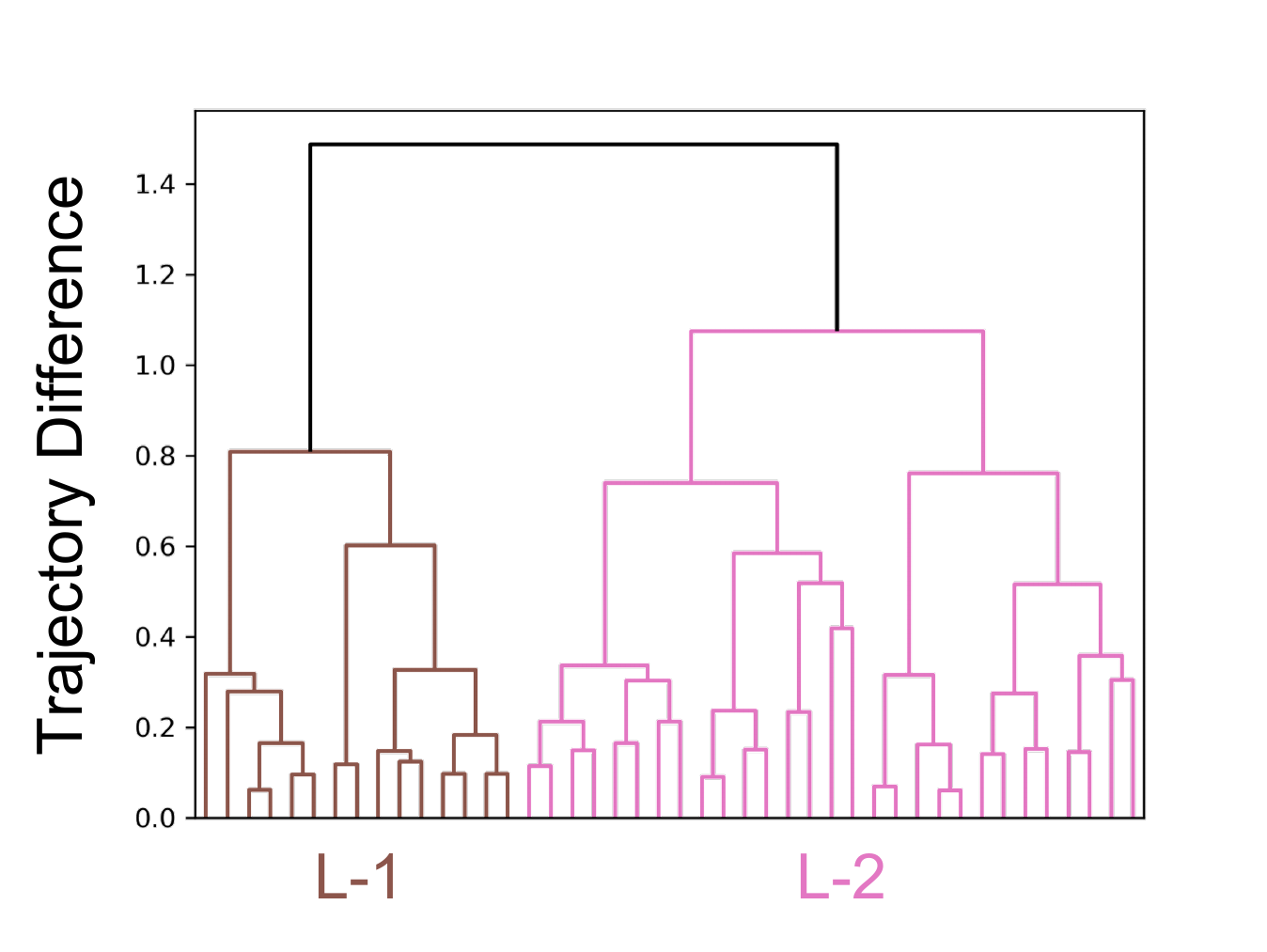}
\caption{\textbf{Hierarchical clustering of 7--6 nucleation trajectories (cutoff = 1.2).} Each dendrogram cluster represents a cluster of nucleation trajectories. The difference between nucleation trajectories is calculated via pairwise distances between normalized histograms of particle path labels (see Materials and Methods). The difference obtained is used in agglomerative hierarchical clustering. With the cutoff of 1.2, two clusters of nucleation trajectories are identified: L-1 with 25 trajectories and L-2 with 156 trajectories.}
\label{fig:ljtraj-hc-2}
\end{figure}

\begin{figure}
\centering
\includegraphics[width=0.7\textwidth]{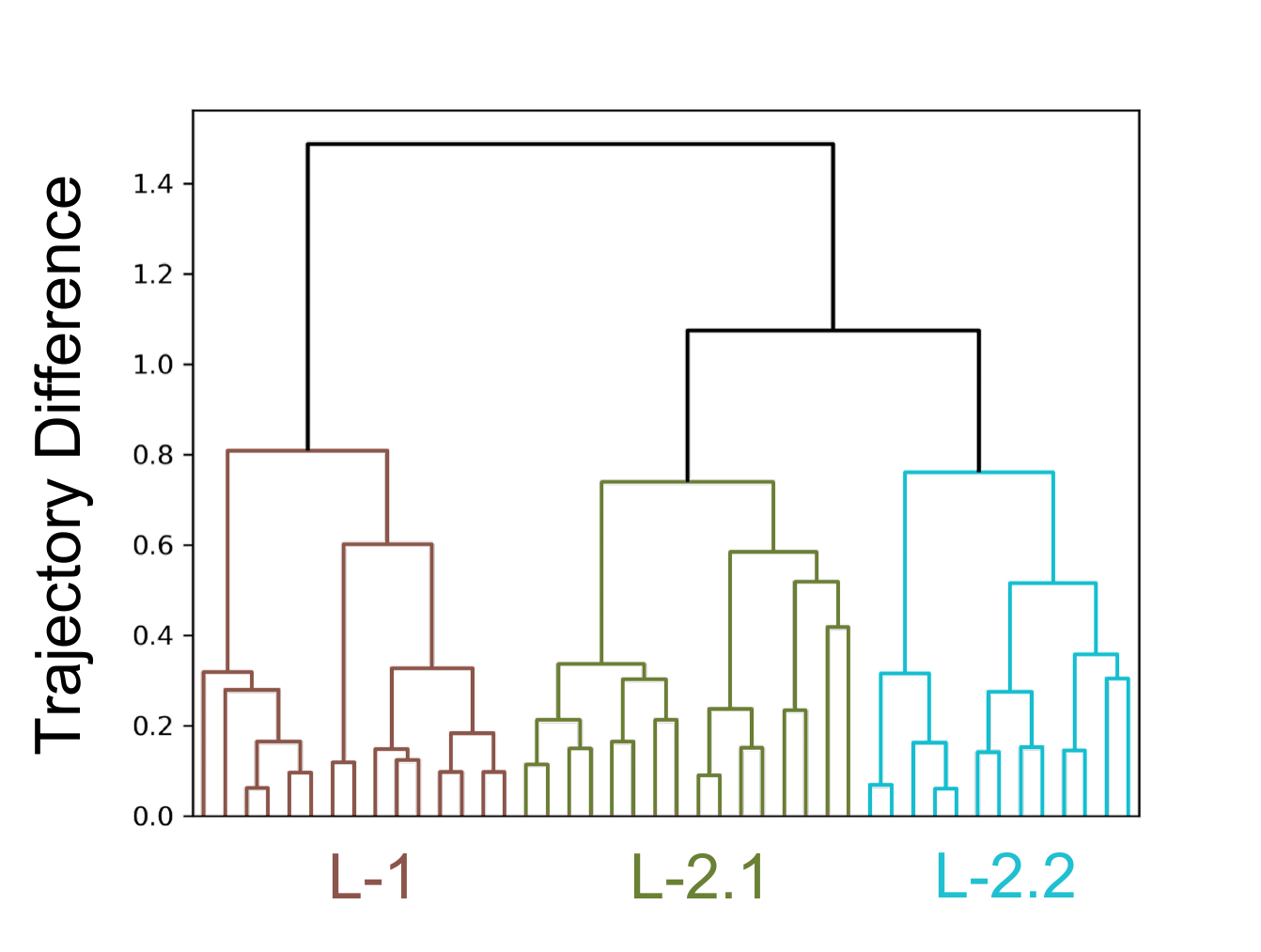}
\caption{\textbf{Hierarchical clustering of 7--6 nucleation trajectories (cutoff = 1.0).} With the cutoff of 1.0, three nucleation trajectory clusters are identified: L-1 with 25 trajectories, L-2.1 with 130 trajectories and L-2.2 with 26 trajectories.}
\label{fig:ljtraj-hc-3}
\end{figure}

\begin{figure}
\centering
\includegraphics[width=0.9\textwidth]{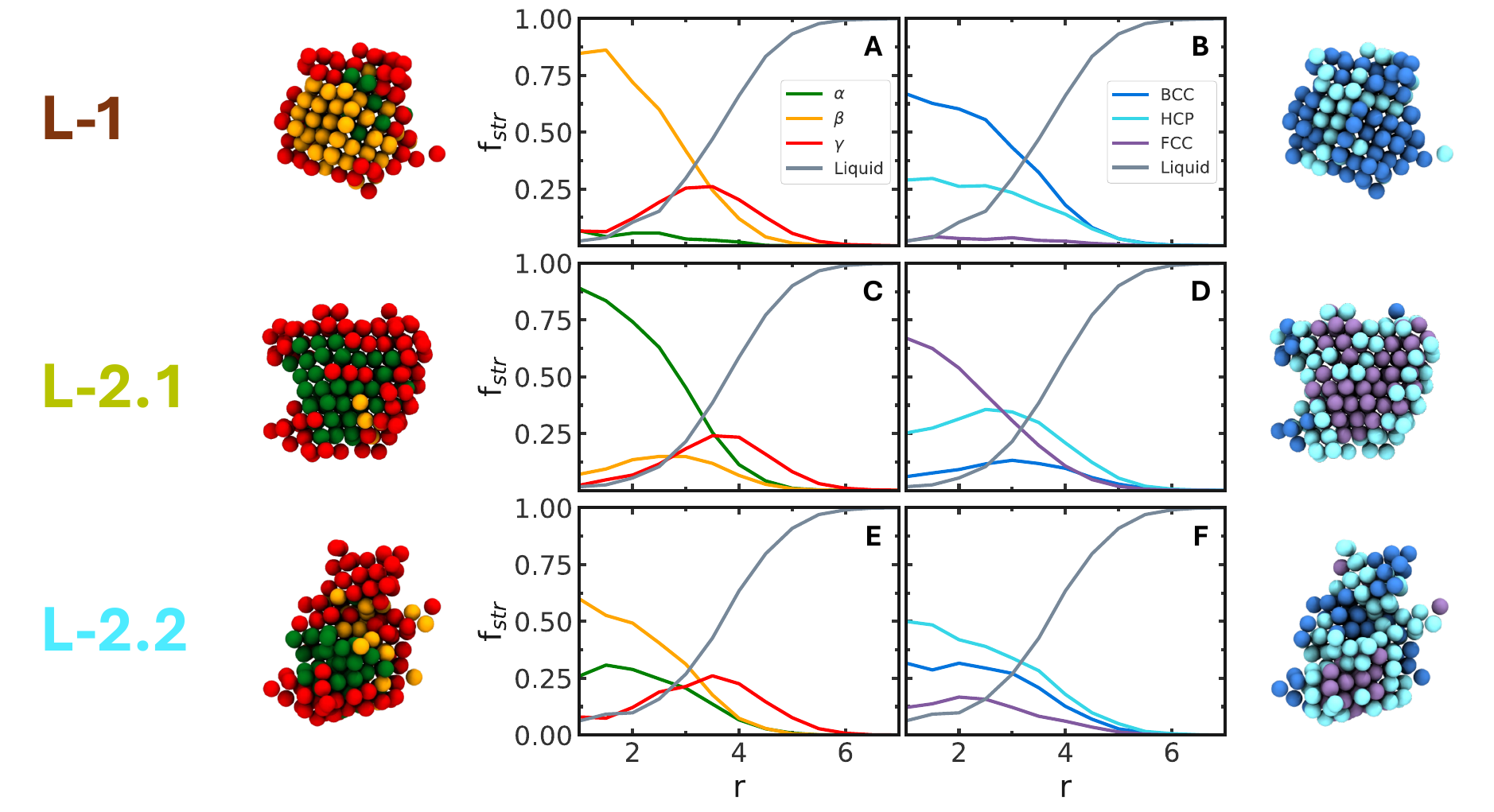}
\caption{\textbf{Radial composition analysis of 7--6 configurations from each trajectory cluster in Fig. \ref{fig:ljtraj-hc-3}.} Panel A, C and E use particle path cluster labels from Fig. 2C for the analysis. Panel B, D and F use the $\overline{q}_4$ and $\overline{q}_6$ values of particles for the analysis. The radial compositions are computed for (A-B) L-1 cluster, (C-D) L-2.1 cluster and (E-F) L-2.2 cluster of trajectories. Configurations with committor estimates of 0.9 or greater are selected for the calculation.}
\label{fig:lj-hc-rca-3}
\end{figure}

\begin{figure}
\centering
\includegraphics[width=0.8\textwidth]{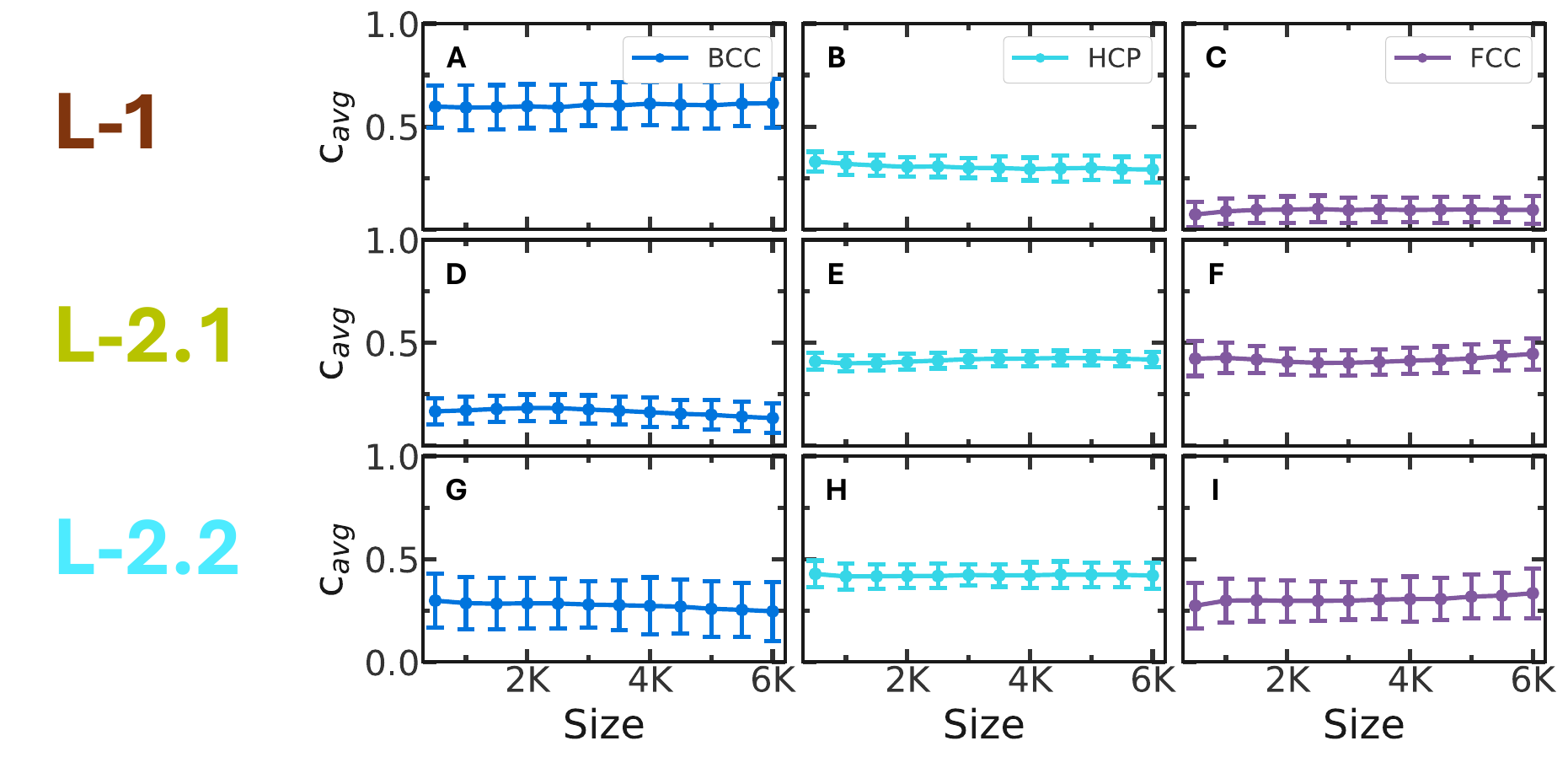}
\caption{\textbf{The average composition of 7--6 nucleation trajectories from each trajectory cluster in Fig. \ref{fig:ljtraj-hc-3}.} The x-axis represents the average nucleus size of the configurations in the trajectories of each respective cluster. c$_\text{avg}$ represents the average composition of the nucleus at a specific size according to the three phases: BCC, HCP and FCC. Blue lines represent the BCC composition, cyan lines represent the HCP composition and purple lines represent the FCC composition. The composition is calculated for (A-C) trajectories in the L-1 cluster, (D-F) trajectories in the L-2.1 cluster, and (G-I) trajectories in the L-2.2 cluster. The error bars reflect the standard deviation of the average composition across all trajectories launched within each trajectory cluster.}
\label{fig:lj-hc-3avgcomp}
\end{figure}

\begin{figure}
\centering
\includegraphics[width=\textwidth]{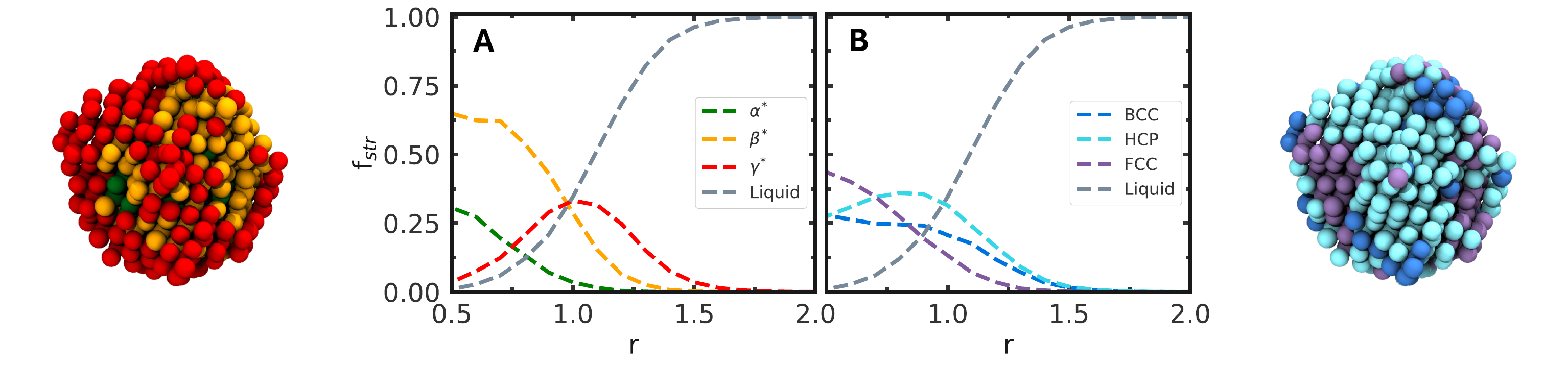}
\caption{\textbf{Radial composition analysis of configurations from Ni$_3$Al nucleation trajectories.} Panel A uses the particle path clusters labels ($\alpha^{*}$, $\beta^{*}$, $\gamma^{*}$) in Fig. 6B for the analysis. Panel B use the $\overline{q}_4$ and $\overline{q}_6$ values of particles for the analysis. Any particles that do not belong to the nucleus are labeled liquid. The analysis is performed using one configuration from each Ni$_3$Al RETIS trajectory with an estimated committor of 0.9 or greater.}
\label{fig:ni3al-rca-all}
\end{figure}

\begin{figure}
\centering
\includegraphics[width=0.7\textwidth]{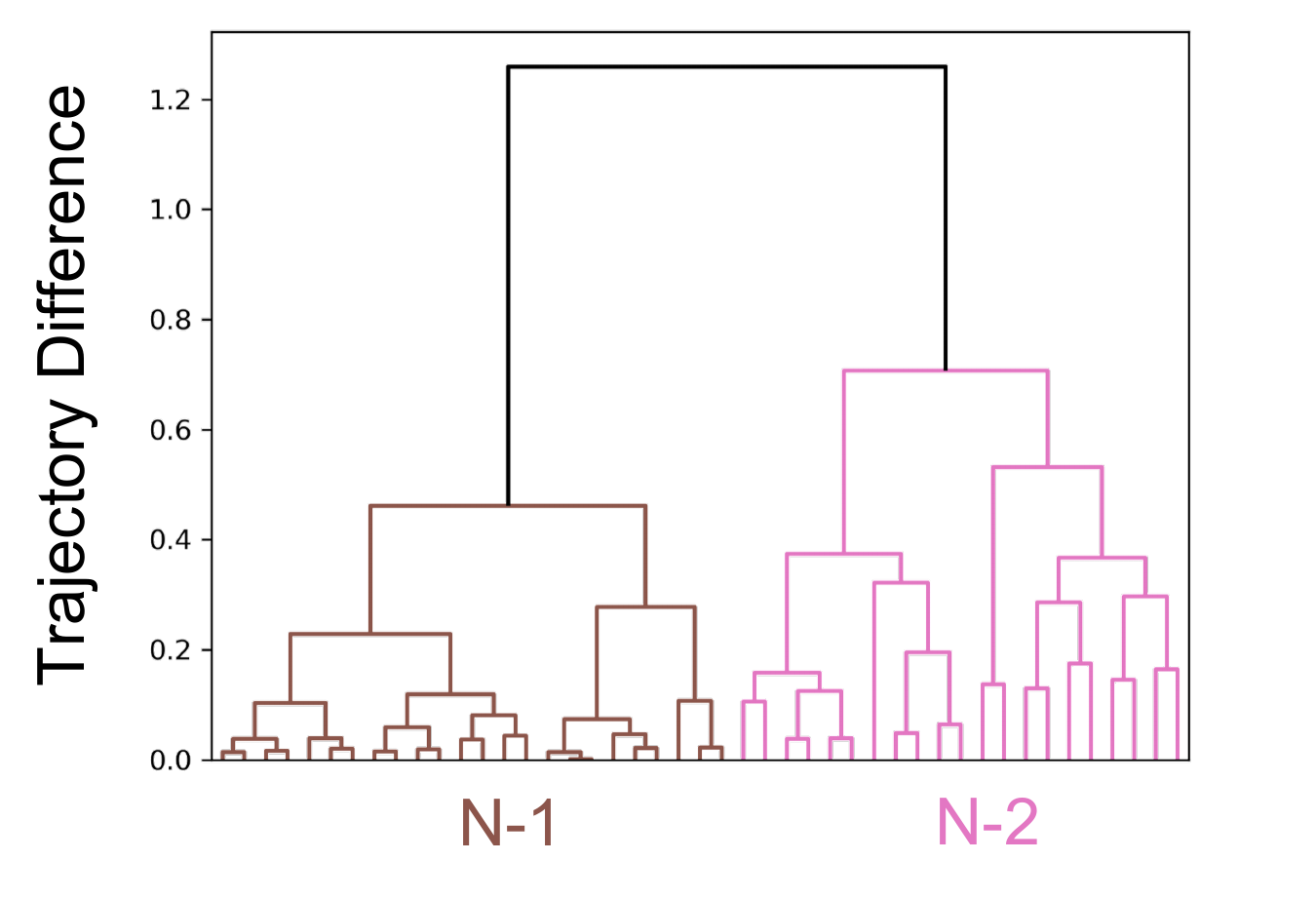}
\caption{\textbf{Hierarchical clustering of Ni$_3$Al nucleation trajectories (cutoff = 0.8).} With the cutoff of 0.8, two nucleation trajectory clusters are identified: N-1 with 45 trajectories and N-2 with 37 trajectories.}
\label{fig:ni3al-hc-2}
\end{figure}

\begin{figure}
\centering
\includegraphics[width=0.7\textwidth]{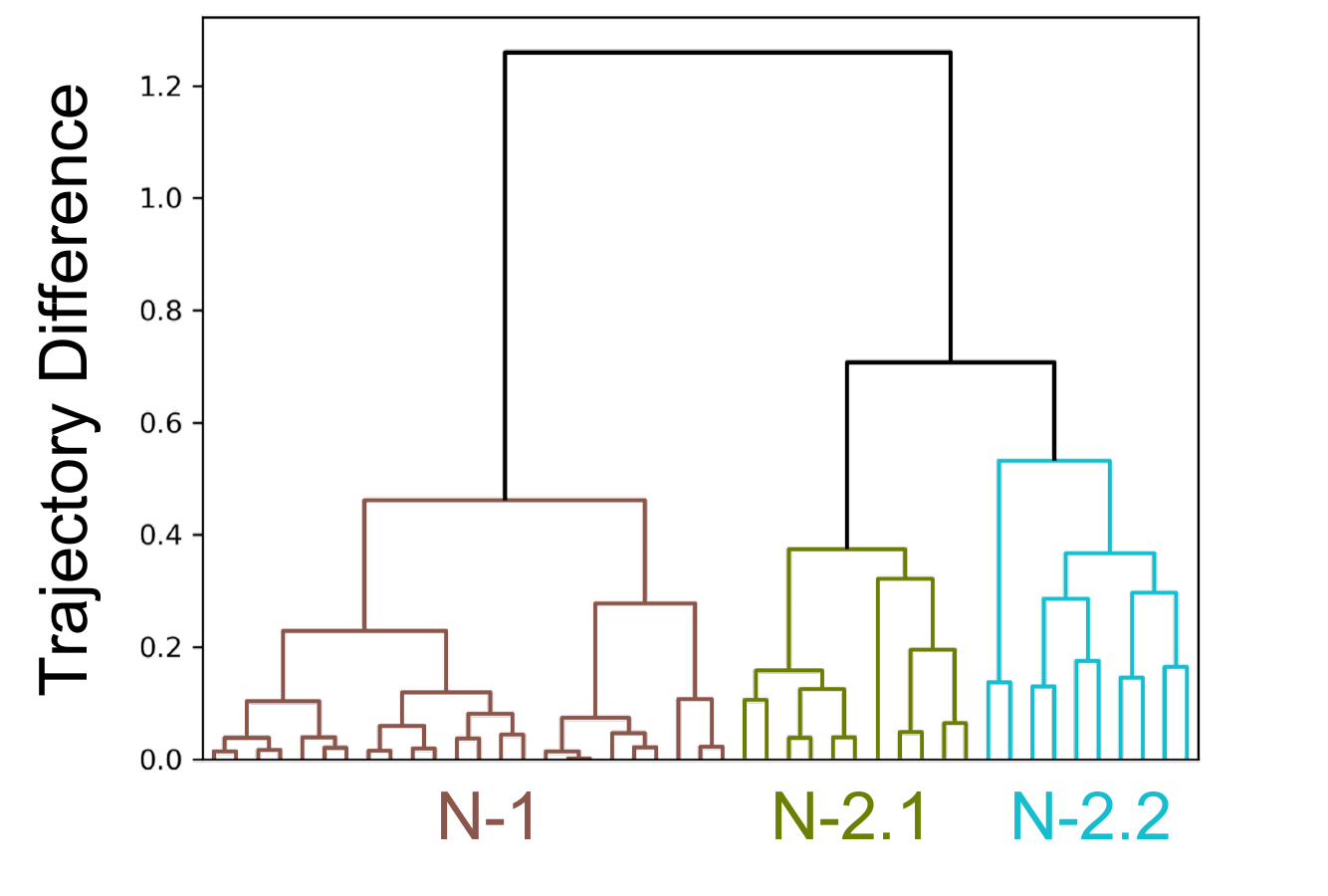}
\caption{\textbf{Hierarchical clustering of Ni$_3$Al nucleation trajectories (cutoff = 0.6).} With the cutoff of 0.6, three nucleation trajectory clusters are identified: N-1 with 45 trajectories, N-2.1 with 13 trajectories and N-2.2 with 24 trajectories.}
\label{fig:ni3al-hc-3}
\end{figure}

\begin{figure}
\centering
\includegraphics[width=\textwidth]{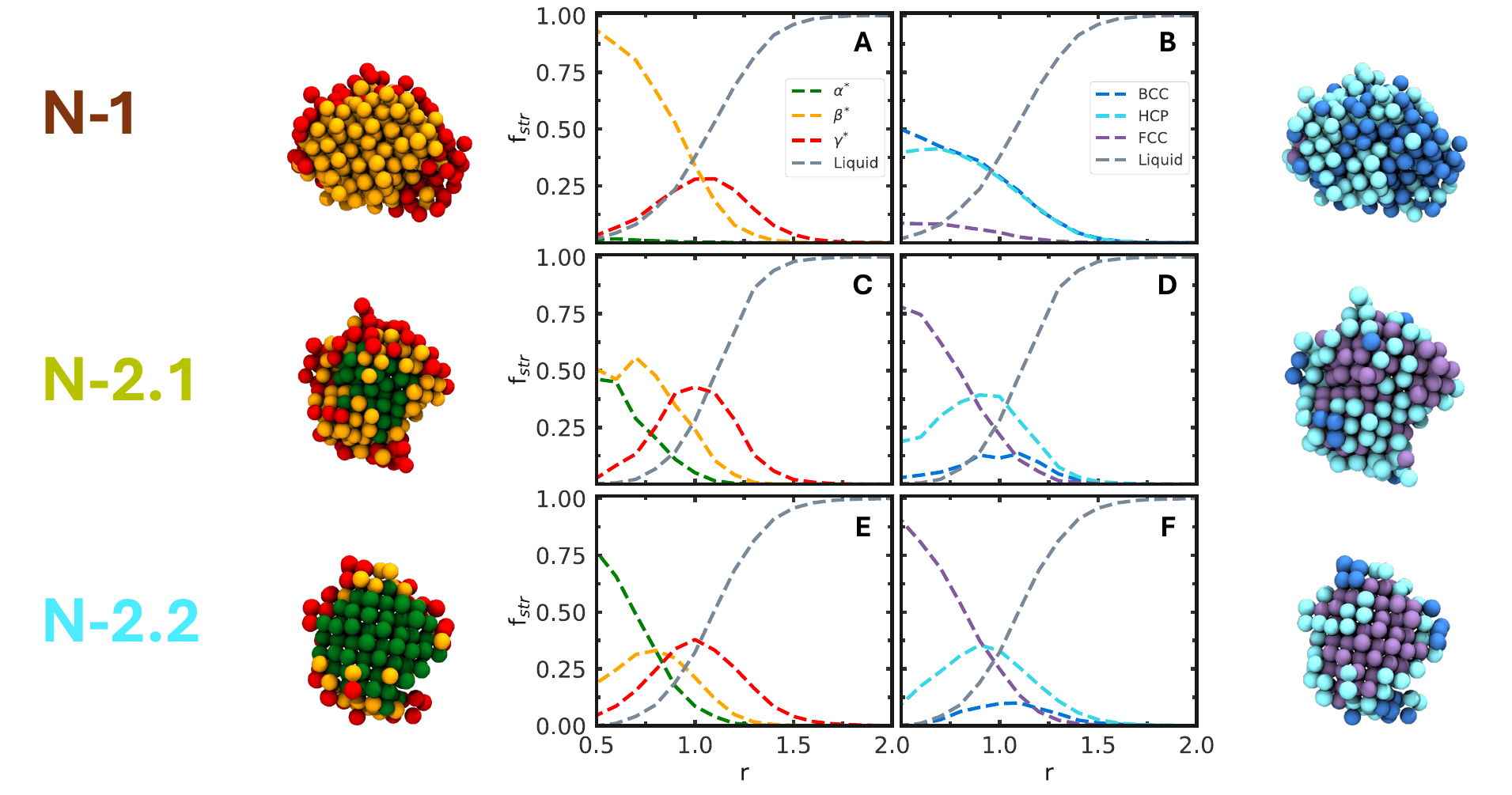}
\caption{\textbf{Radial composition analysis of Ni$_3$Al configurations from each trajectory cluster in Fig. \ref{fig:ni3al-hc-3}.} The radial compositions are computed for (A-B) N-1 cluster, (C-D) N-2.1 cluster and (E-F) N-2.2 cluster from Fig. \ref{fig:ni3al-hc-3}. Panel A, C and E use the particle path cluster labels from Fig. 6B. Panel B, D and F label the particles according to $\overline{q}_4$ and $\overline{q}_6$ values. Configurations with committor estimates of 0.9 or greater are selected for the calculation.}
\label{fig:ni3al-hc-rca-3}
\end{figure}

\begin{figure}
    \centering
    \includegraphics[width=0.8\textwidth]{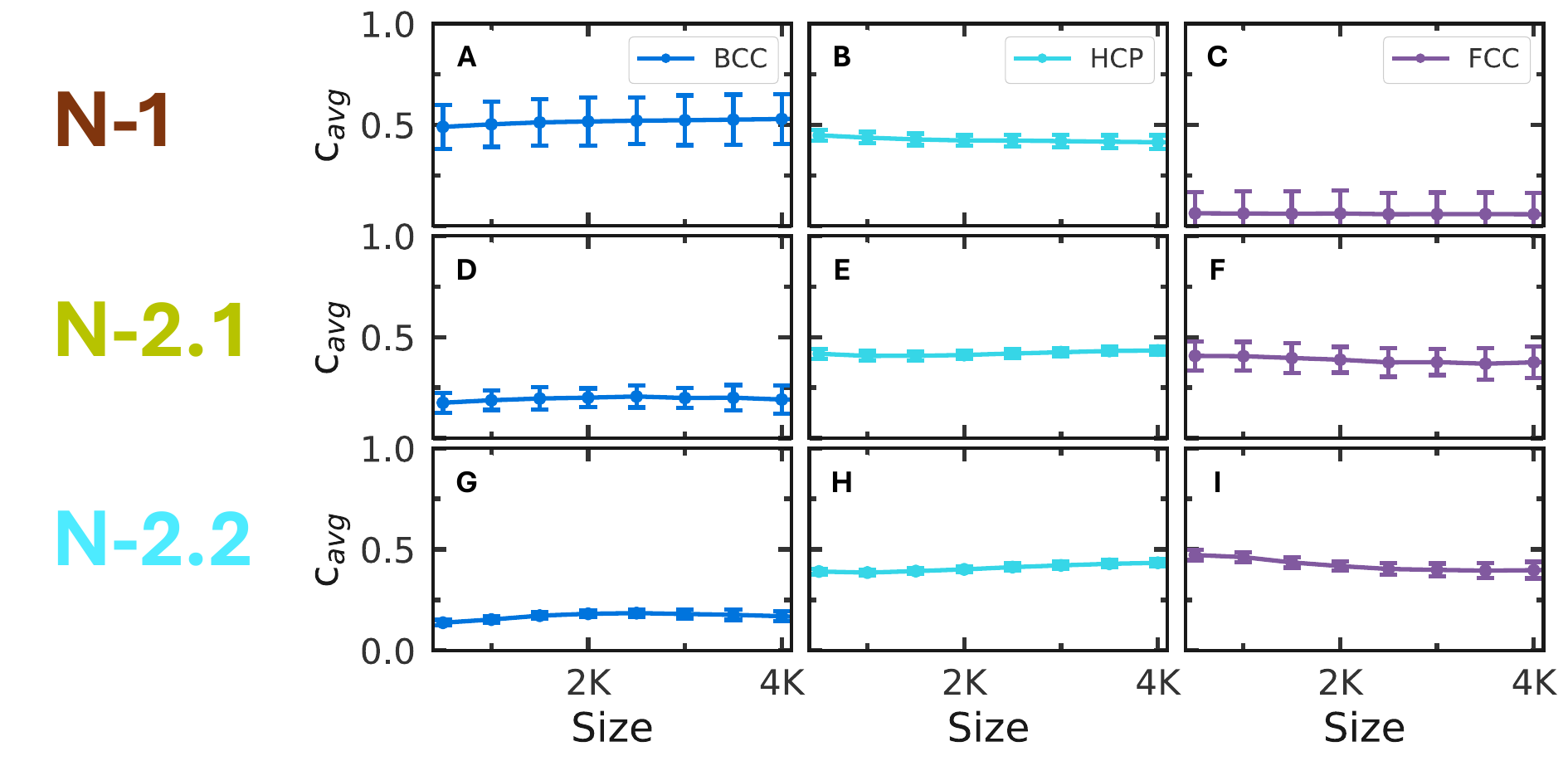}
    \caption{\textbf{Average composition of Ni$_3$Al nucleation trajectories from each trajectory cluster in Fig. \ref{fig:ni3al-hc-3}.} The composition is calculated for (A-C) trajectories in the N-1 cluster, (D-F) trajectories in the N-2.1 cluster and (G-I) trajectories in the N-2.2 cluster. The error bars reflect the standard deviation of the average composition across all trajectories launched within each trajectory cluster.}
    \label{fig:ni3al-hc-3avgcomp}
\end{figure}